%% file: main.tex
\documentclass[conference]{IEEEtran}
\IEEEoverridecommandlockouts
\usepackage{graphicx}
\usepackage{siunitx}
\usepackage{amssymb}
\usepackage{algorithm}  
\usepackage{amsmath,amssymb,amsfonts}
\usepackage{algorithmicx}  
\usepackage{algpseudocode} 
\usepackage{xcolor}
\usepackage{enumitem}
\usepackage[backend=biber]{biblatex}
\def\BibTeX{{\rm B\kern-.05em{\sc i\kern-.025em b}\kern-.08em
		T\kern-.1667em\lower.7ex\hbox{E}\kern-.125emX}}

\newcommand{\Fig}[1]{Figure~\ref{#1}}

\newcommand{\Tbl}[1]{Table~\ref{#1}}
\newcommand{\Sec}[1]{Section~\ref{#1}}

\title{DLFusion: An Auto-Tuning Compiler for Layer Fusion on Deep Neural Network Accelerator}
\author{
	\IEEEauthorblockN{1\textsuperscript{st} Zihan Liu}
	\textit{Shanghai Jiao Tong University}\\
	\and
	\IEEEauthorblockN{2\textsuperscript{nd} Jingwen Leng}
	\textit{Shanghai Jiao Tong University}\\
	\and
	\IEEEauthorblockN{3\textsuperscript{rd} Quan Chen}
	\textit{Shanghai Jiao Tong University}\\
	\and
	\IEEEauthorblockN{4\textsuperscript{th} Chao Li}
	\textit{Shanghai Jiao Tong University}\\
	\and
	\IEEEauthorblockN{5\textsuperscript{th} Wenli Zheng}
	\textit{Shanghai Jiao Tong University}\\
	\and
	\IEEEauthorblockN{6\textsuperscript{th} Li Li}
	\textit{Shanghai Jiao Tong University}\\
	\and
	\IEEEauthorblockN{7\textsuperscript{th} Minyi Guo}
	\textit{Shanghai Jiao Tong University}\\
}
\usepackage[backend=biber]{biblatex}
\begin{document}
	\maketitle
	\input{"./Abstract.tex"}
	\input{"./Introduction.tex"}

\input{"./Background.tex"}

	\input{"./Motivation.tex"}
	\input{"./Implementation.tex"}
	\input{"./Evaluation.tex"}

\input{"./related.tex"}

	\input{"./Conclusion.tex"}
	\input{"./Reference.tex"}
\end{document}

%% file: Abstract.tex
\begin{abstract}
Many hardware vendors have introduced specialized deep neural networks (DNN) accelerators owing to their superior performance and efficiency.
As such, how to generate and optimize the code for the hardware accelerator becomes an important yet less explored problem.
In this paper, we perform the compiler-stage optimization study using a novel and representative Cambricon DNN accelerator and demonstrate that the code optimization knobs play an important role in unleashing the potential of hardware computational horsepower.
However, even only two studied code optimization knobs, namely the number of cores and layer fusion scheme, present an enormous search space that prevents the naive brute-force search.
This work introduces a joint, auto-tuning optimization framework to address this challenge.
We first use a set of synthesized DNN layers to study the interplay between the hardware performance and layer characteristics.
Based on the insights, we extract the operation count and feature map channel size as each layer's characteristics and derive a joint optimization strategy to decide the performance-optimal core number and fusion scheme.
We evaluate the performance of the proposed approach using a set of representative DNN models and show that it achieves the minimal of 3.6x and the maximal of 7.9x performance speedup compared to no optimization baseline.
We also show that the achieved speedup is close to the oracle case that is based on a reduced brute-force search but with much less search time. 
\end{abstract}


%% file: Introduction.tex
\section{Introduction}

Deep learning has achieved great success  in the key application domains such as computer vision and natural-language processing.
The derived deep neural network (DNN) models have significant requirement for computation and memory resources, which exceed the capability of the existing architectures. 
Owing to the repeated common computation pattern in different DNN models, such as 2D convolution and matrix multiplication, both the academia and industry begin to embrace the specialized hardware accelerators~\cite{PuDianNao,TPU,V100} for their high performance and energy efficiency. 
Compared to the general-purpose architecture like CPU/GPU, hardware accelerators are specialized for a specific task and have simplified control logic so that they can dedicate more resources for computation and memory structure~\cite{PuDianNao}.
Their design decision has also lead to their distinctive programming models from the CPU/GPU.

General-purpose architectures have well-defined ISAs so that the compiler can perform various performance optimizations.
However, optimizing the code for the hardware accelerators is challenging because hardware vendors do not expose the accelerator's low-level ISA.
Instead, hardware vendors provide a SDK with high-level API for application/model developers, but the SDK is highly abstracted and difficult to control the exact hardware behavior.
As such, significant efforts are devoted in computational-graph level optimization such as operator fusion, operator concatenation~\cite{TVM}, and operator substitutions~\cite{TASO}.
Nonetheless, the graph-level optimization are independent of underlying hardware, cannot be used to tune the performance for a specific accelerator.


In this paper, we explore the code optimization for a novel accelerator Cambricon MLU100~\cite{CambriconMLU100}. MLU100 has a higher peak performance in FP16/INT8 than Tesla V100~\cite{V100}, but also requires highly optimized code to fully unleash its computational horsepower.
The accelerator SDK includes both high-level and low-level APIs.
The high-level APIs are highly abstracted and have little optimization space. 
In contrast, the low-level API exposes two execution hyper-parameters, which are number of cores and layer fusion schemes.
Our characterization results show that we need to carefully and wisely select those two parameters to achieve the optimal performance for a specific DNN model. 
However, even with only two parameters, the search space is too large for a brute-force search.

\begin{figure}
	\centering
	\includegraphics[width=0.9\linewidth]{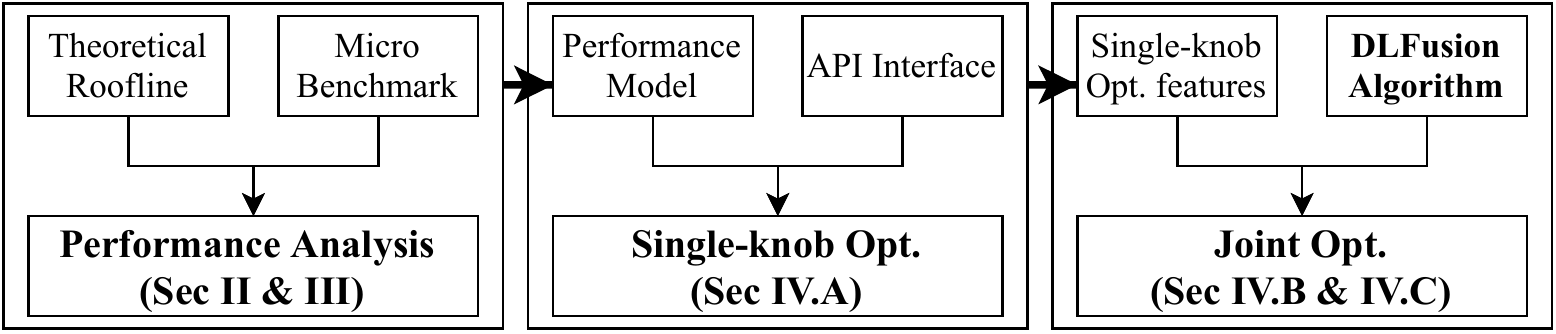}
	\caption{\label{fig-arch}Overall flow of the optimizer design in our work.}
\end{figure}

On the basis of the low-level SDK, we propose an compiler-stage auto-tuning optimizer, \emph{DLFusion}, for MLU100 accelerator which performs joint optimization for the two exposed execution hyper-parameters. 
To the best of our knowledge, this work is the first to consider arbitrary auto-fusion patterns that are mathematically equivalent.
In contrast, existing fusion optimizations such as TensorRT and XLA~\cite{TensorRT, TensorFlowXLA} are rule-based and therefore can only support a limited set of pre-defined fusion patterns.
The basic architecture of the optimizer is shown in \Fig{fig-arch}, and this paper is organized as followed.
In \Sec{sec-BGD}, we first show the performance gap between theoretical model and actual execution using a series of micro-benchmark, then we analyze the optimization knobs and performance in detailed in \Sec{sec:motivation}, by which we present our performance model for later optimization. Moreover, we find it difficult to achieve the optimal settings for different networks by btute methods, so an auto-tunning optimization is necessary. 
In \Sec{sec-Impl}.A, we first propose single knob optimization for Model Parallelism using the aforementioned performance model, then we integrate fusion scheme into the optimization procedure guided by the insight of \Sec{sec-Impl}.B, finally, we present our optimization algorithm and software implementation in \Sec{sec-Impl}.C.

%% file: Background.tex
\section{Investigated Platform}\label{sec-BGD}

In this paper, we use Cambricon MLU-100-C3~\cite{CambriconMLU100}, a dedicated deep learning accelerator designed by Cambricon Technologies. 
The MLU-100 can deliver the computing performance with much better energy efficiency than general-purpose processors like CPU and GPU. 
In this section, we first give a brief introduction of the accelerator and then conduct an analysis to reveal the factors that can impact the accelerator's performance efficiency.

\subsection{Experimental Setup} 

\subsubsection{Hardware Setup}
\Tbl{tab-HWSpec} lists the detailed specifications of the studied Cambricon MLU100 accelerator.
The accelerator has 32 cores in total, where each core has the computation power 1 TFLOPS in FP32, 2 TFLOPS in FP16, and 4 TFLOPS in INT8. 
The accelerator can support common computation patterns in deep neural network models, including the convolution layer, ReLU layer, and batch normalization layer~\cite{CambriconMLU100}.
However, its core microarchitecture is not revealed, which we use a microbenchmark-based methodology to study.
With those auto-generated microbenchmarks covering different computational intensity and operation count, we can quickly have a high-level understanding of the target hardware's computational characteristics. 
One of the salient features of our work is that, for other accelerators with different microarchitectures, this microbenchmark methodology can also be applied to reveal hardware characteristics for optimization.

		\begin{table}[h]
			\caption{The MLU100 hardware specification.}
			\centering
			\begin{tabular}{|c|c|}
				\hline
				Item & Descriptions \\
				\hline
				Core freq. & 1GHz \\
				Float perf. (FP16) & 64 TFLOPS \\
				Integer perf. (INT8) & 128 TOPS \\
				Memory size & 8 GB\\
				Memory bandwidth & 102.4 GB/s\\
				Memory bit width & 256-bit\\
				Host Interface & PCIe 3.0x16\\
				TDP & 110 W \\
				ECC Enabled & Yes\\
				\hline
			\end{tabular}
			\vspace*{0.2cm}
			\label{tab-HWSpec}
		\end{table}
		\begin{figure}[h]
			\centering
			\includegraphics[width=\linewidth]{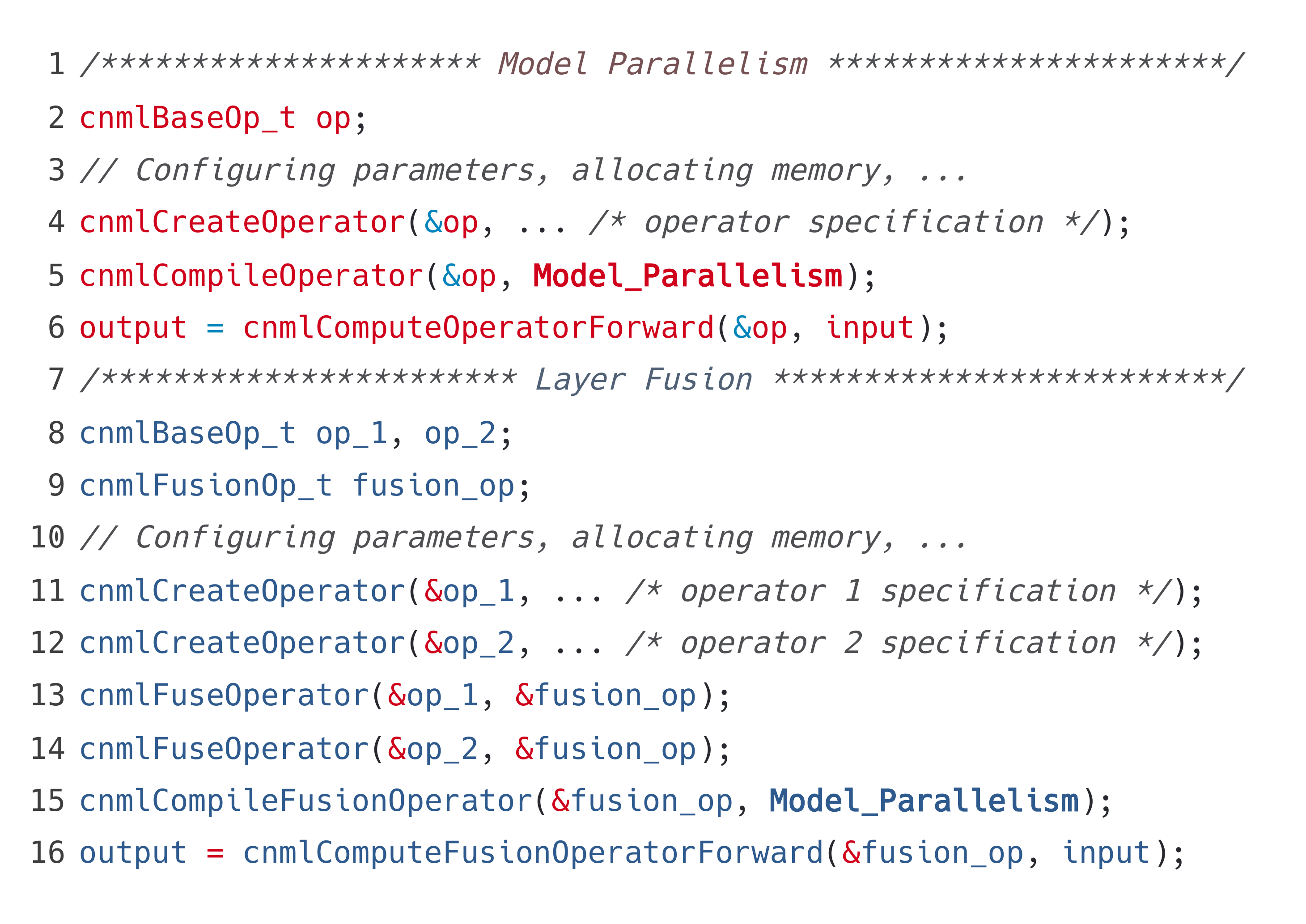}
			\caption{SDK code sample for setting MP and layer fusion.}
			\label{fig-CodeSeg}
		\end{figure}

\subsubsection{Software Setup}
The chip vendor provides an operator-level SDK and a high-level runtime library for programming the accelerator. 
The operator-level SDK, called CNML, supports common operators such as convolution, ReLU, and BatchNorm, which are used in computer vision and natural language processing models.
The CNML supports two hyper-parameters for running those operators, as shown in \Fig{fig-CodeSeg}.
The first hyper-parameters is \textit{model parallelism} (\textit{MP}), which specifies the number of cores (up to 32) used by the operator.
The second hyper-parameter is \textit{layer fusion}, which specifies the number of layers that are fused for concurrent execution and therefore increased parallelism. 
These two hyper-parameters can represent the execution of fusion operation on multi-core architecture based accelerators, and this optimization is orthogonal to other graph-level optimizations including Common Subexpression Elimination (CSE)~\cite{TensorFlowXLA}, operator substitution~\cite{TASO}, etc.
In this work, we use this operator-level SDK and tune the hyper-parameter settings to optimize different DNN models.


\subsection{Performance Analysis}

We first construct a single-layer based microbenchmark to study the accelerator's performance efficiency for DNN model layers with different characteristics. 
We focus on the convolutional layer (\texttt{Conv}) and fully connected layer (\texttt{FC}) because they represent most of the computation in today's DNN models~\cite{TC}.
For each layer, we sweep its different parameters and compute their required operation count through Equation~\ref{equ-GOP-CONV} and~\ref{equ-GOP-FC} respectively.
We first perform the experiment using the single core and then multiple cores to understand the impact of the core number on performance efficiency.

\begin{equation}
GOPS_{Conv} \gets 2\times H_{Out} * W_{Out} * H_{K} * W_{K} * C_{In} * C_{Out}
\label{equ-GOP-CONV}
\end{equation}
\begin{equation}
GOPS_{FC} \gets 2 \times M \times K \times N
\label{equ-GOP-FC}
\end{equation}
\subsubsection{Single-core Performance}
We first use the simple roofline model~\cite{roofline} to model the performance of convolution and fully-connect layers under different parameters, and the operation intensity is calculated as equation~\ref{equ-int}. 
\begin{equation}
Intensity \gets \frac{GOPS}{\sum(sizeof(tensors))}
\label{equ-int}
\end{equation}

\begin{figure}[h]
	\centering
	\includegraphics[width=0.8\linewidth]{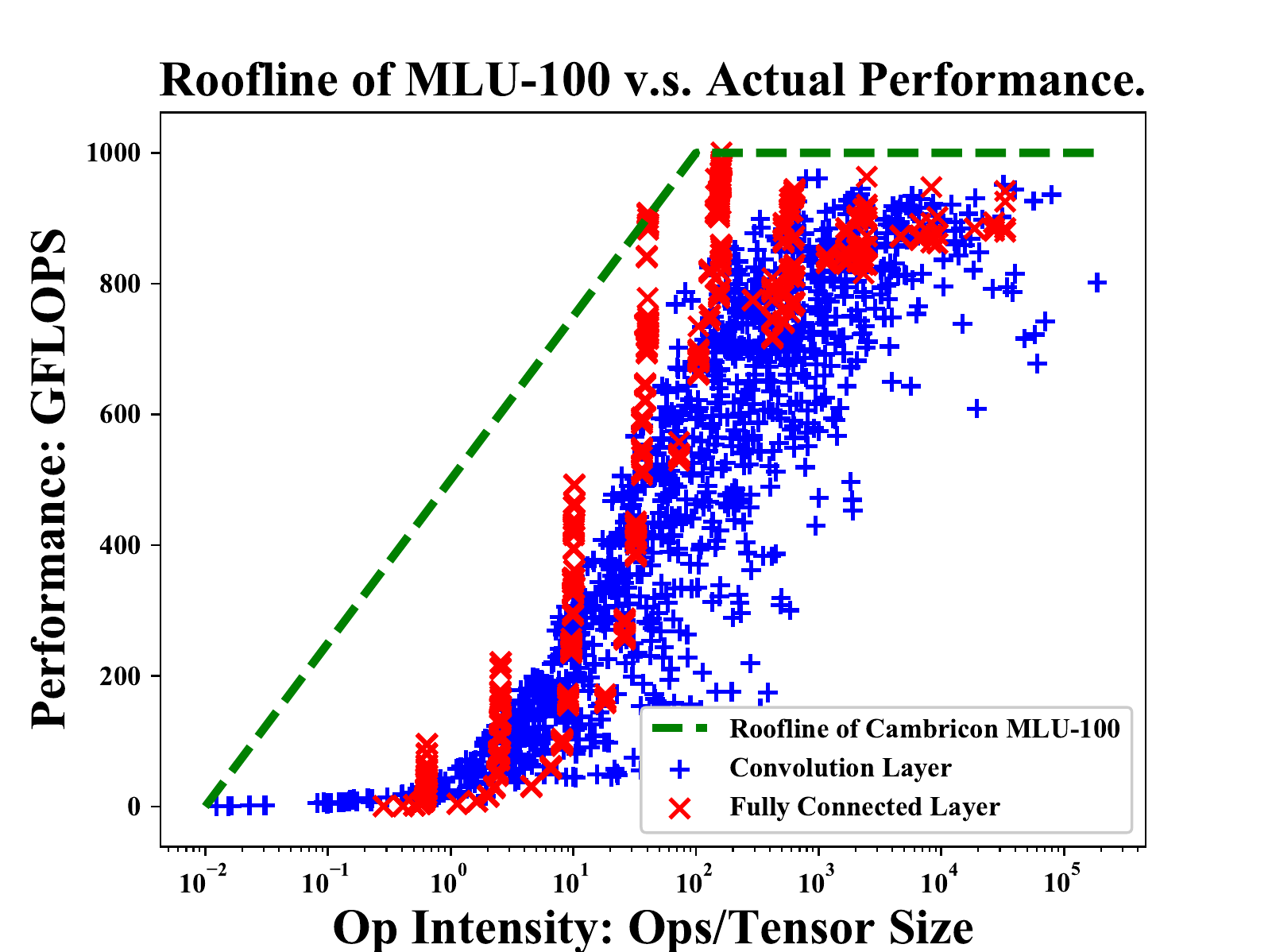}
	\caption{\label{fig-rfline} Roofline model of Cambricon MLU-100 and actual performance. }
\end{figure}
However, as shown in \Fig{fig-rfline}, there's significant gap between the exact performance and theoritical performance on Cambricon MLU-100, moreover, operation intensity in roofline model can not distinct the performance of operations under the same intensity effectively leading to the difficulty in modeling the performance.

So, we applied \texttt{PCA} method to extract the parameters that are most likely to influence the performance (by which we can adjust the hardware resource assigned to the operation and thus improve the performance), and for Cambricon MLU-100, we found that operation count has the most significant influence on the performance, and channel the second.

\begin{figure*}[htbp]
	\centering
	\includegraphics[width=0.32\linewidth]{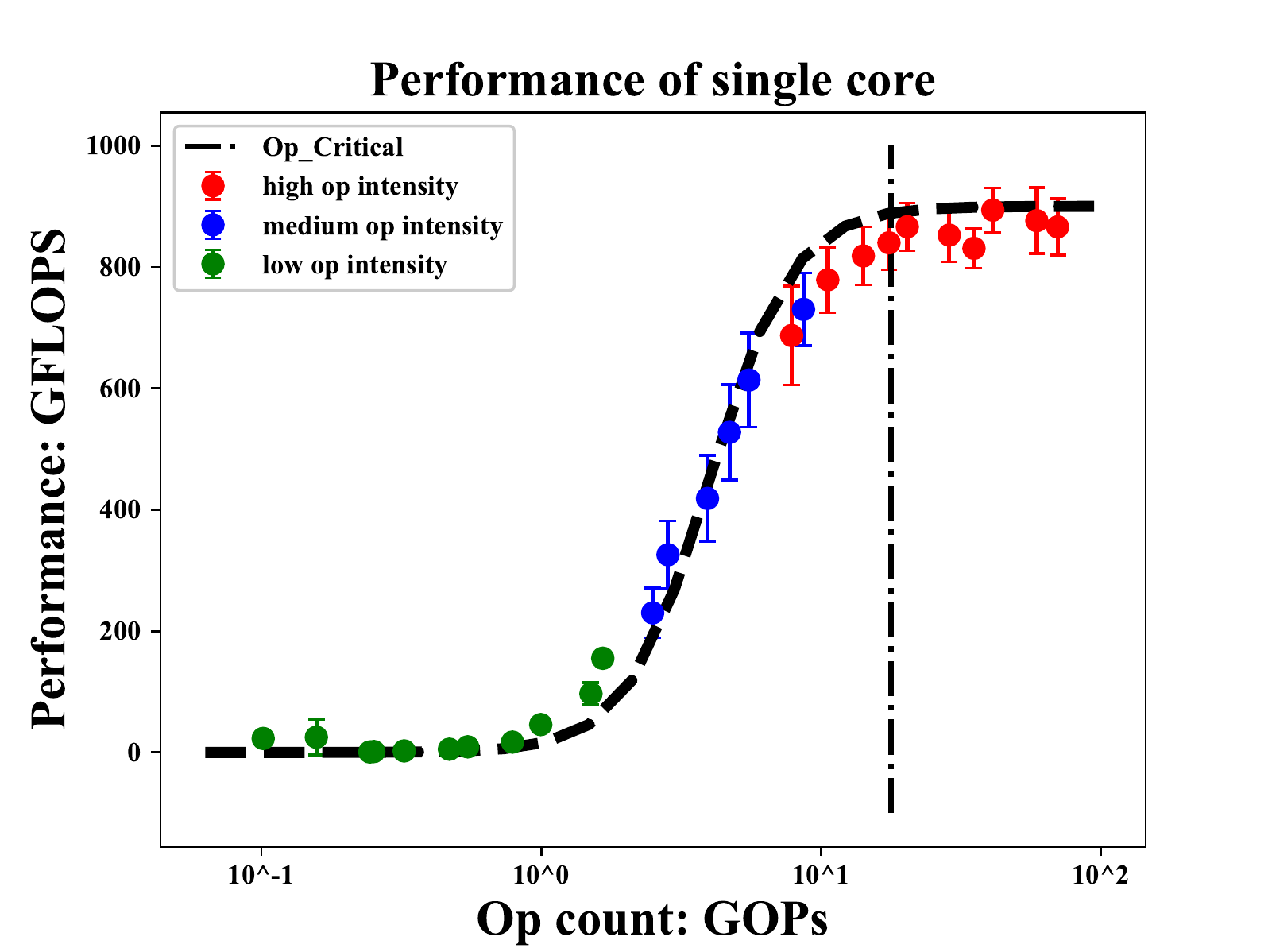}
	\includegraphics[width=0.32\linewidth]{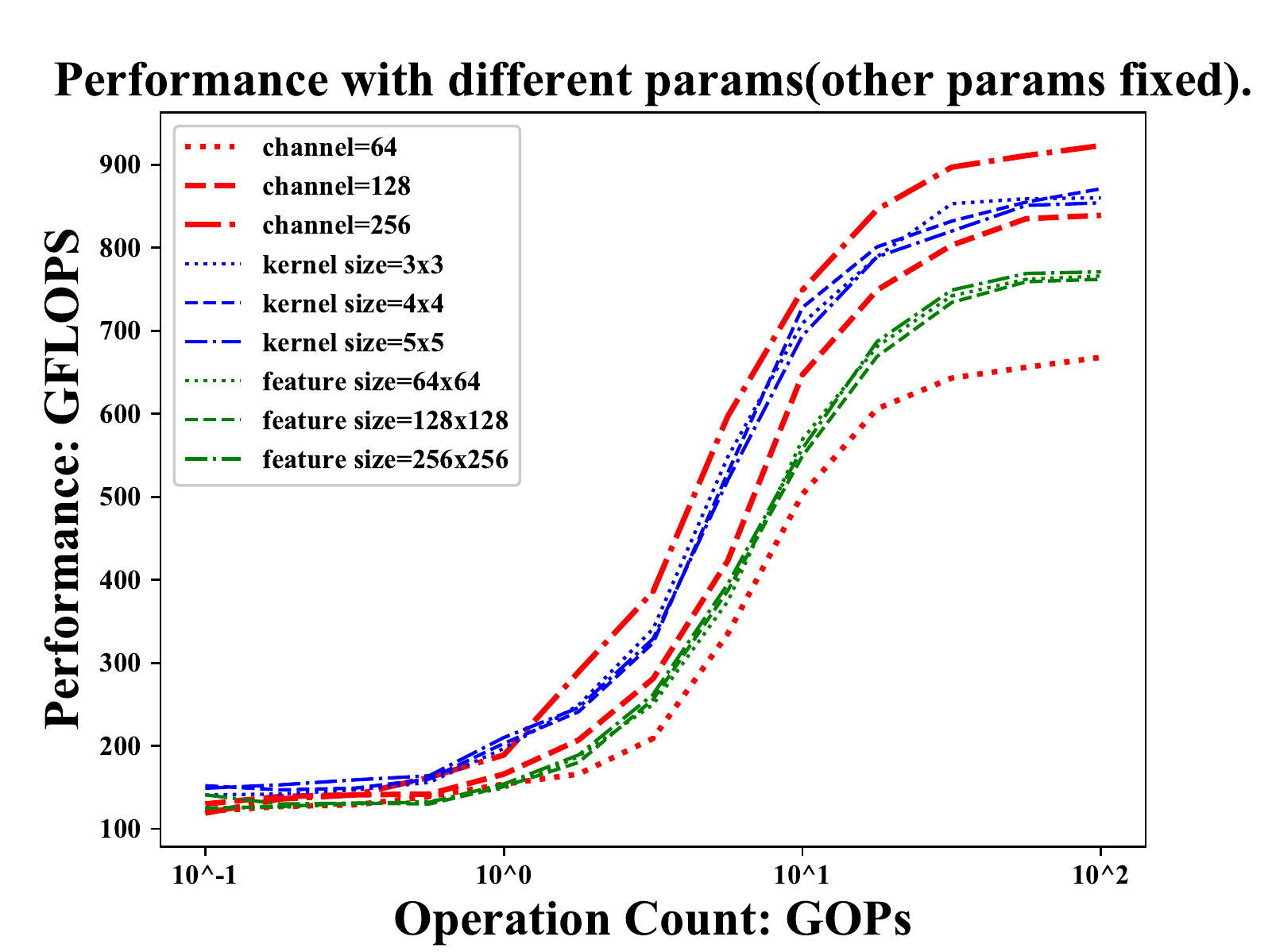}
	\includegraphics[width=0.32\linewidth]{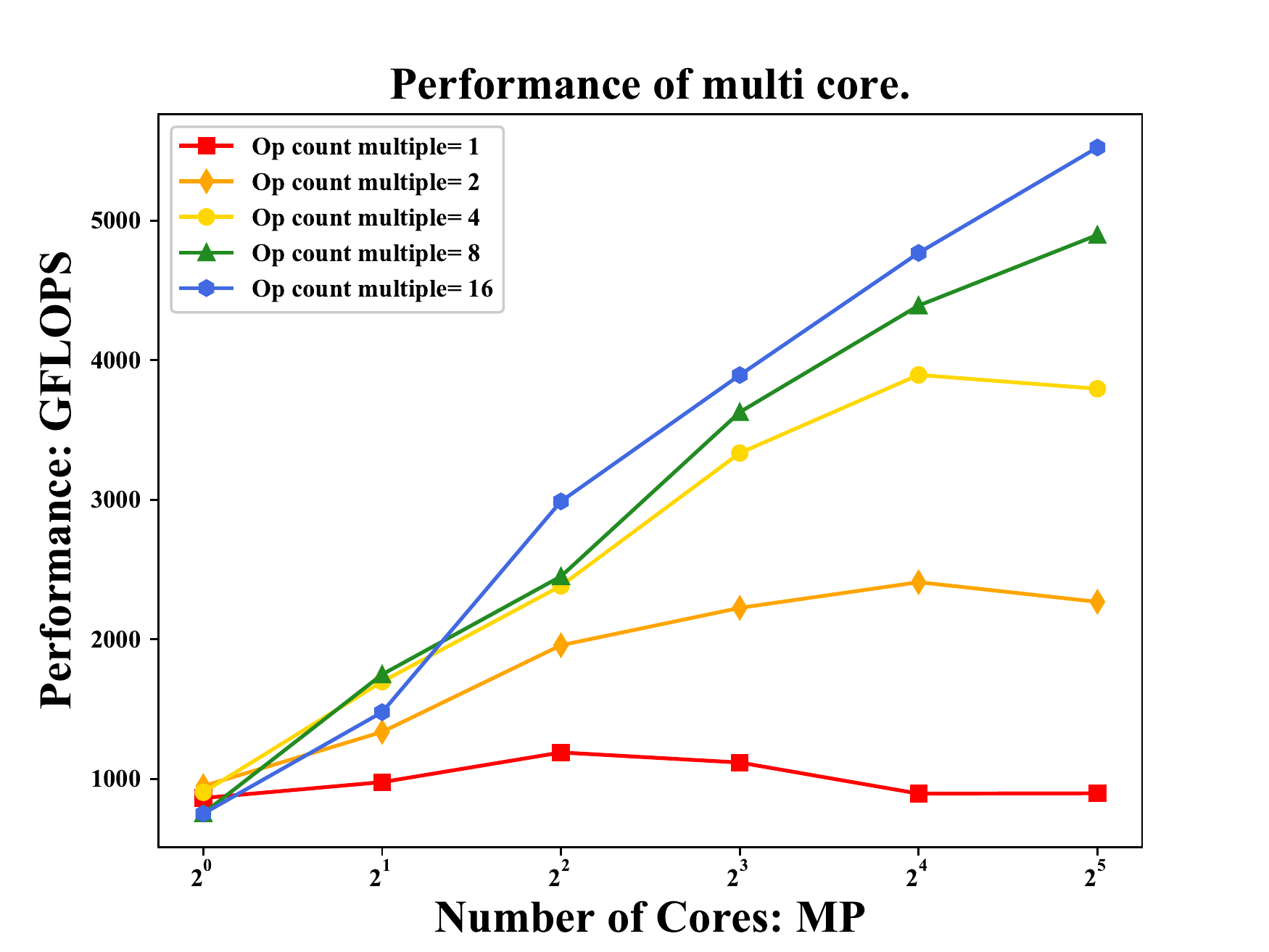}
	\caption{\label{fig-profiling} (a) Single core performance of ops. (b) Influence convolution parameter with other parameters fixed. (c) Multi core performance of ops.}
\end{figure*}

\Fig{fig-profiling}(a) shows the relationship between the layer's operation count and achieved performance efficiency measured by GFLOPS (Giga floating-point operations per second), and the operations with similar op count have similar performance (the error bar stands for the standard diviation in the figure is short).
As \Fig{fig-profiling} shows, layers performance efficiency is largely determined by its operation count: the higher the operation count, the better performance efficiency on the accelerator, and once the operation count reaches a critical value, the performance will not increase. We called it $ OpCount_{critical} $ and will be used in our optimization process. 

On the other hand, layers with medium and high intensity exhibit a larger variability of performance efficiency for the layers with the same operation count, according to the PCA result, channel should be the main reason. To verify the assumption, we observe the performance of convolution layers with different channel/kernel size/output image size with other parameters fixed.
As shown in \Fig{fig-profiling}(b), We found that channel have non-negligible influence on layer's performance. Actually, the errorbar in \Fig{fig-profiling}(a) is mainly introduced by changing channel. 
For kernel size and feature size, they contribute little to distinct the performance of different layers, which match the result of PCA well.
As such, we also explore the setting of the parameters with first and second largest influence according to PCA result: channel of convolution.

\subsubsection{Multi-core Performance}

The above single-core performance experiment shows that the operation count  of a layer impacts the accelerator performance efficiency. 
Based on this observation, we further study the impact of the number of cores  with varying operation count.
For this experiment, we start with a fixed convolutional layer from the VGG-19 model~\cite{VGG}: \texttt{Input/Output Channel=64}, \texttt{Output Size=$ 224\times 224 $}, \texttt{Kernel Size=$ 3\times 3 $}, for which we use the notation of $\{64, 64, 224\times224, 3\times3\}$ to represent its parameters in the following sections.
We increase the operation count of the layer via expanding the \texttt{Channel} dimension by different factors.
\Fig{fig-profiling}(c) shows that the layers with large operation count prefer a large number of cores. Layers with small (moderate) operation count prefer a small (moderate) core numbers to achieve the best performance.


%% file: Motivation.tex
\section{Performance Analysis of Optimization Knobs}\label{sec:motivation}

After analyzing the performance features of the DNN accelerator, we focus on the problem of achieving the best performance for a give DNN model, i.e., lowest inference latency. This section motivates the need for an auto-tuning compiler by demonstrating that the hyper-parameter setting for achieving the best performance for a given DNN model is highly dependent on model characteristics.

\begin{figure}[h]
	\centering
	
	\includegraphics[width=0.8\linewidth]{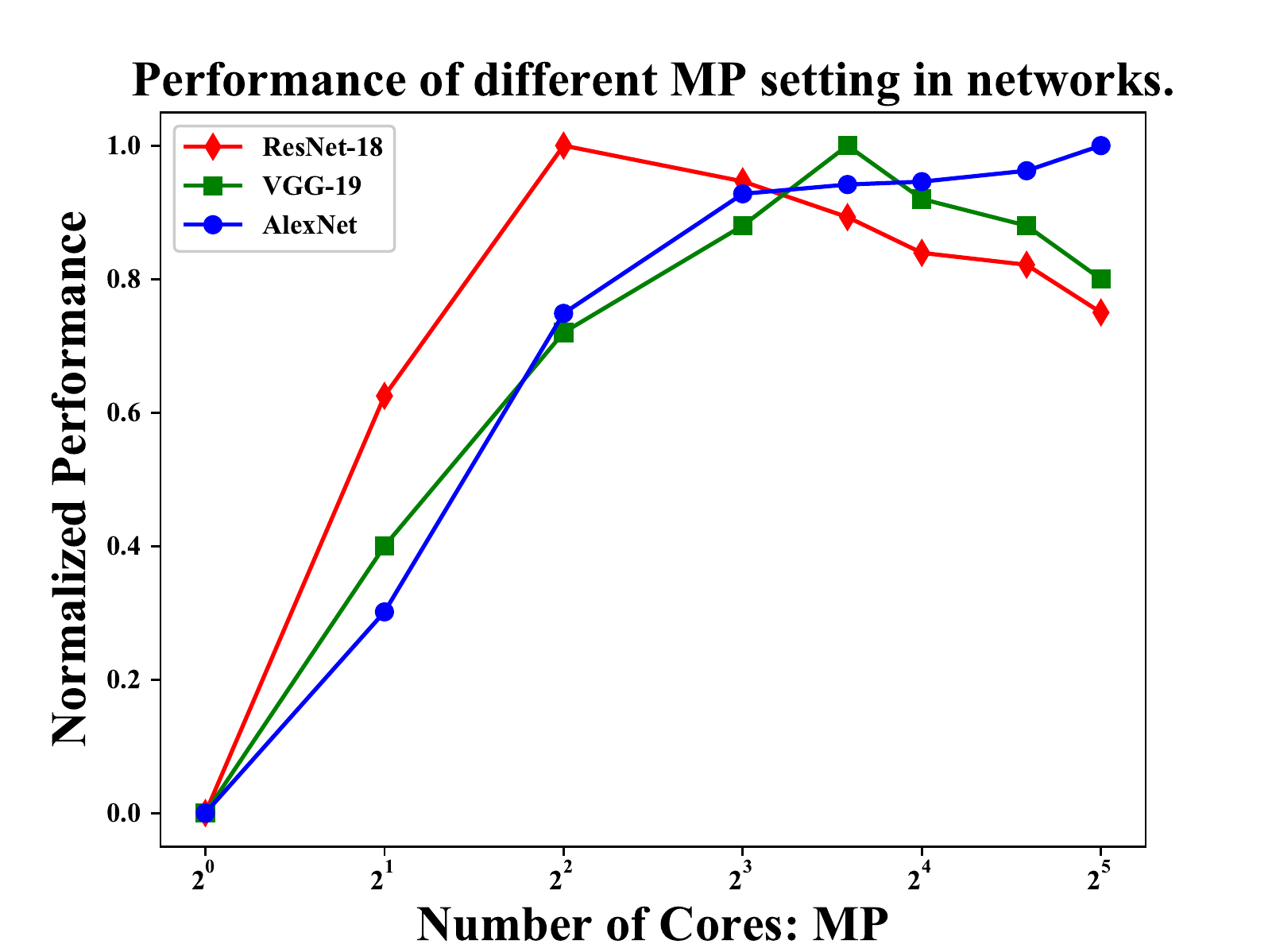}\\
	\includegraphics[width=0.8\linewidth]{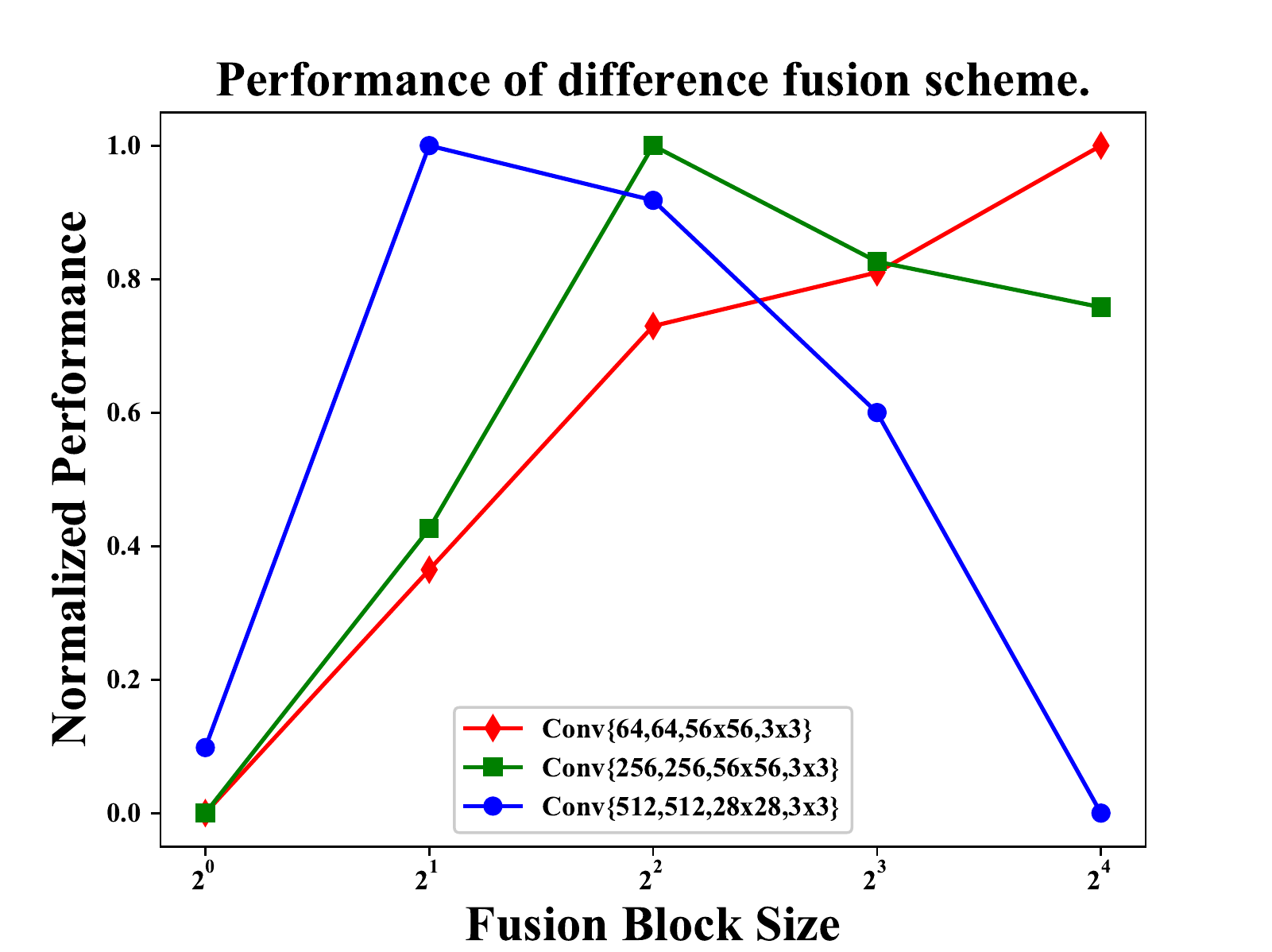}

	\caption{\label{fig-Motivation} (a) Optimal MP setting of different networks with all layers sharing the same MP. (b) Optimal fusion block size of convolutions with different parameters.}

\end{figure}

In this work, we focus on the aforementioned two hyper-parameters, namely \texttt{model parallelism} and \texttt{layer fusion} for optimizing the performance of a DNN model on the MLU-100 accelerator. For the convenience of the experiment, we first study the two hyper-parameters separately.

\subsection{Model Parallelism.}

As we describe in \Sec{sec-BGD}, \texttt{model parallelism} ($MP$) represents the number of cores for running the given DNN layer.
We sweep this $MP$ hyper-parameter for different DNN models and show the results in \Fig{fig-Motivation}(a).
The results show that \textbf{using the maximum number of cores does not necessarily lead to the best performance}.
The optimal core number for ResNet-18 and VGG-19 is 4 and 16, respectively.
The reason is that when the $MP$ is too large, each core is dispatched with less number of operation count, leading to net performance degradation. As such, it is essential to find a method to set the optimal $MP$ based on the characteristic of different DNN models.


\subsection{Layer Fusion.}

This layer fusion hyper-parameter lets users combine multiple layers into a single block, which has two benefits.
It first increases the concurrent operation count than the layer-wise, no-fusion execution.
Consider an example of fusing two layers: the computation of the second layer can start when the first layer's output is partially available. 
It also reduces the cost of off-chip memory round trip because the output of a layer can be generated on-chip and immediately reused.

Prior work like TVM~\cite{TVM} only considers the fusion of convolutional layers and other types of layers such as ReLu and batch normalization. However, the vendor-provided CNML programming frameworks supports the fusion of almost arbitrary types and numbers of layers.
One of the major differences between our work and TVM is that we consider multiple convolution layers in a fusion block.
As a result, the optimization space is much larger and therefore more challenging.
 

To study its impact on the performance, we construct three CNN models, each of which has 16 identical baseline \texttt{Conv} layers.
The three baseline layers, which are selected from ResNet~\cite{ResNet} and VGG~\cite{VGG}, have parameters of  $ \{64, 64, 56\times56,3\times3\} $, $\{256, 256, 56\times56, 3\times3\}$, and $\{512, 512, 28\times28, 3\times3\}$, respectively. 

In this experiment, we sweep the fusion block size $B_{size}$ for executing each CNN model, which leads to $16/B_{size}$ fused blocks.
As \Fig{fig-Motivation} shows, different models have different optimal fusion block sizes.
The reason is that although layer fusion has two major benefits, it also involves redundant computation that needs a careful trade-off. If the fusion block size is too large, the redundant calculation will even degrade the overall performance.  
As such, we must also find a strategy to decide the optimal fusion scheme for different CNN models.



\subsection{Infeasibility of Brute-Force Search.}

For the optimal performance, we must jointly consider the fusion scheme and model parallelism, which leads to a huge space that makes the brute-force search infeasible.
Assume a CNN model with $n$ layers, Equation~\ref{equ-Space} gives the number of possible combinations for setting the two hyper-parameters.
When $n$ equals 50, there are $ 8.17\times10^{75} $ possible combinations.
Based on our experimental insights, we propose an intelligent auto-tuning optimizer that uses the inherent characteristics of CNN models to quickly identify their hyper-parameter setting.  
\begin{equation}
\label{equ-Space}
Space(n)=\sum_{i=1}^{n-1}(32^{i+1}\times \frac{\prod_{x=1}^{i}(n-x)}{i!})
\end{equation}

%% file: Implementation.tex
\section{DLFusion Approach}\label{sec-Impl}
In this section, we present the design and implementation of \emph{DLFusion}, which jointly optimizes the two knobs, \textit{model parallelism, MP} and \textit{layer fusion}, to maximize the inference performance of DNN models on the target MLU-100 accelerator. 
We first explain how to select the optimal \texttt{MP} for a given DNN layer, and then explain how to determine the layer fusion scheme for multiple layers based on their characteristics. 
In the end, we formalize our optimization algorithm and present its implementation details.
 
\begin{figure}[t]
	\centering
	\includegraphics[width=0.8\linewidth]{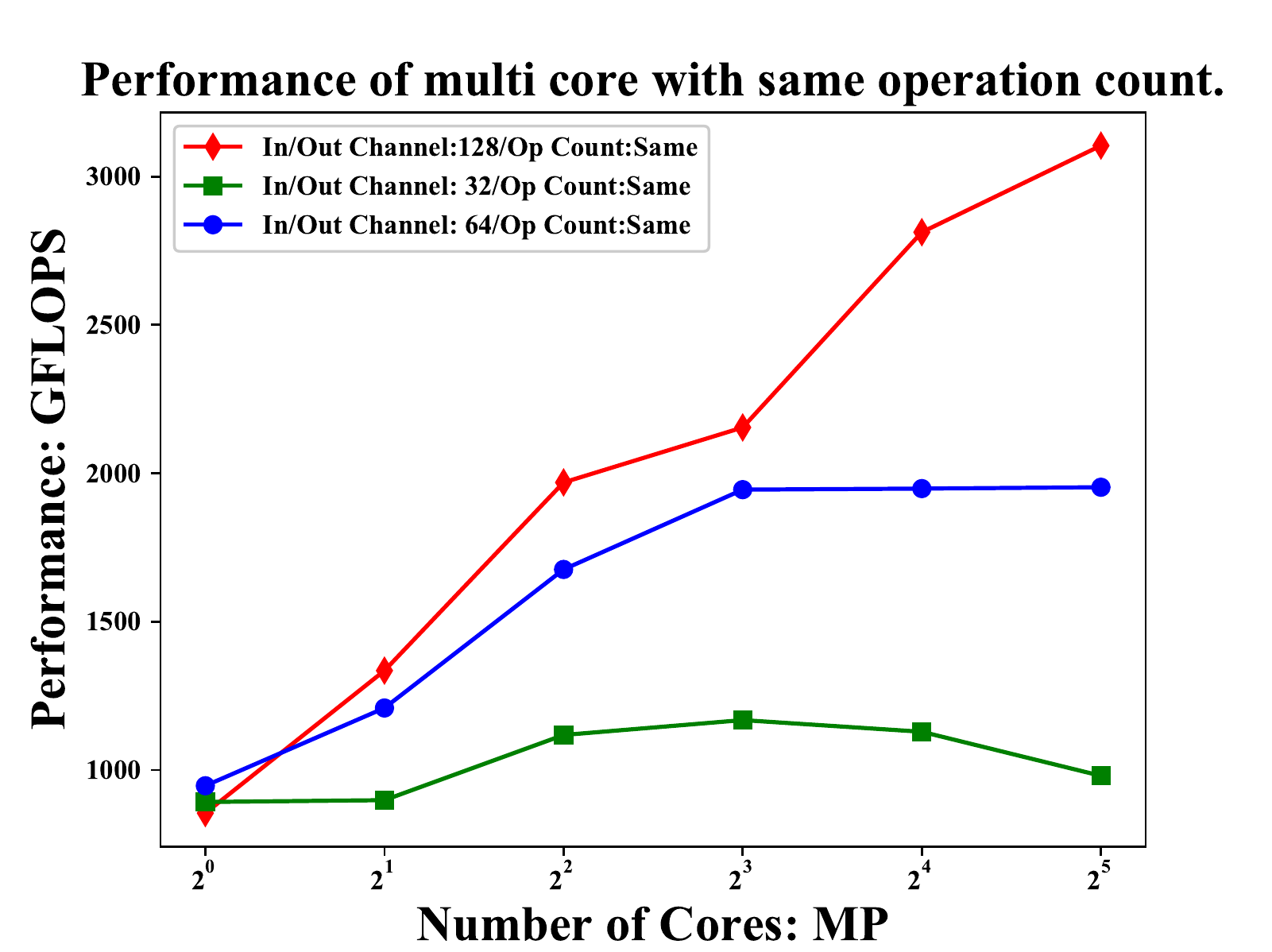}\\
	\includegraphics[width=0.8\linewidth]{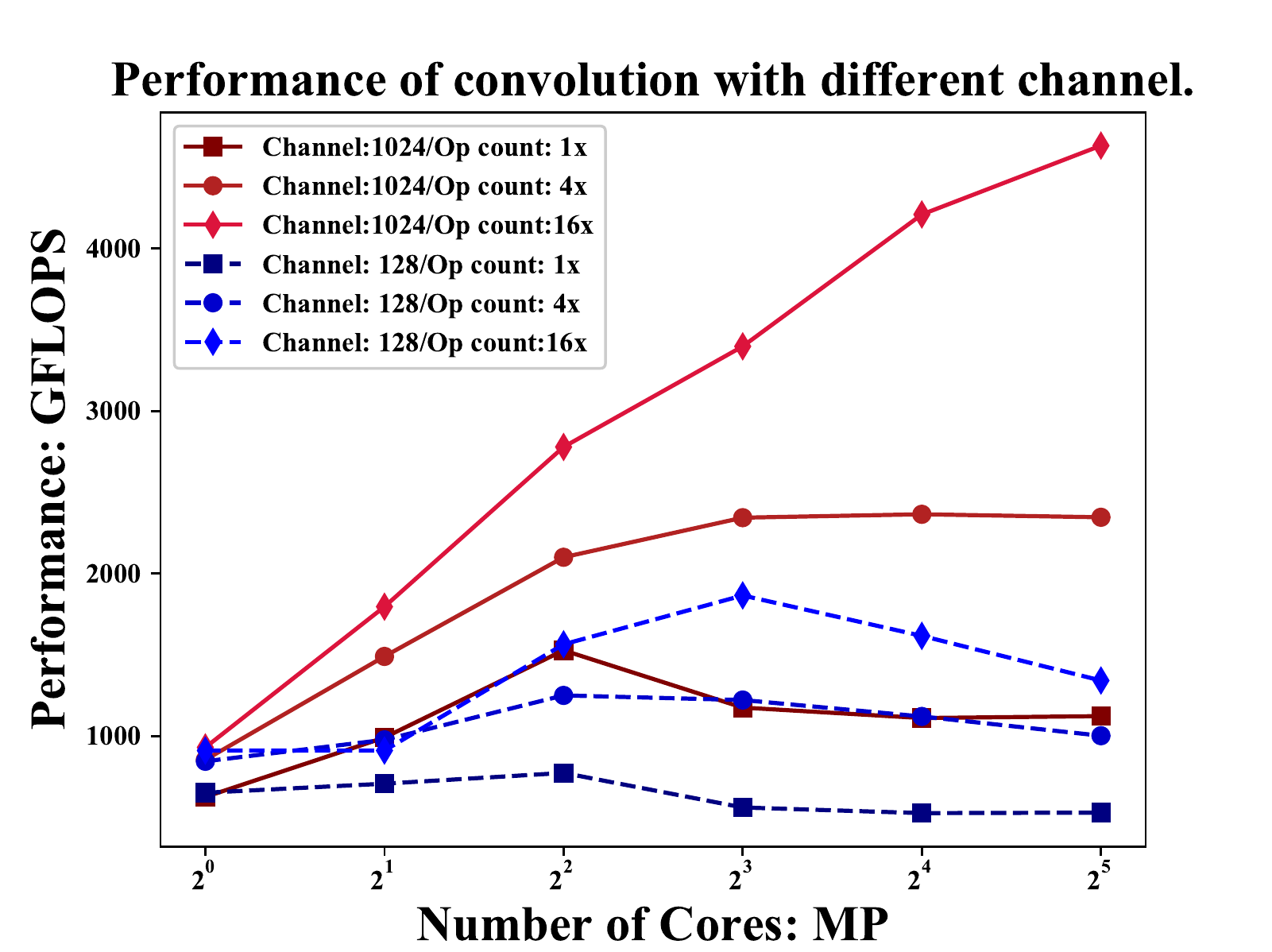}
	
	\caption{\label{fig-SMCPerfDetail} (a) Multi-core performance fixing operation count. (b) Multi-core performance fixing channel.}
\end{figure}
\begin{figure*}[htbp]
	\centering
	\includegraphics[width=0.32\linewidth]{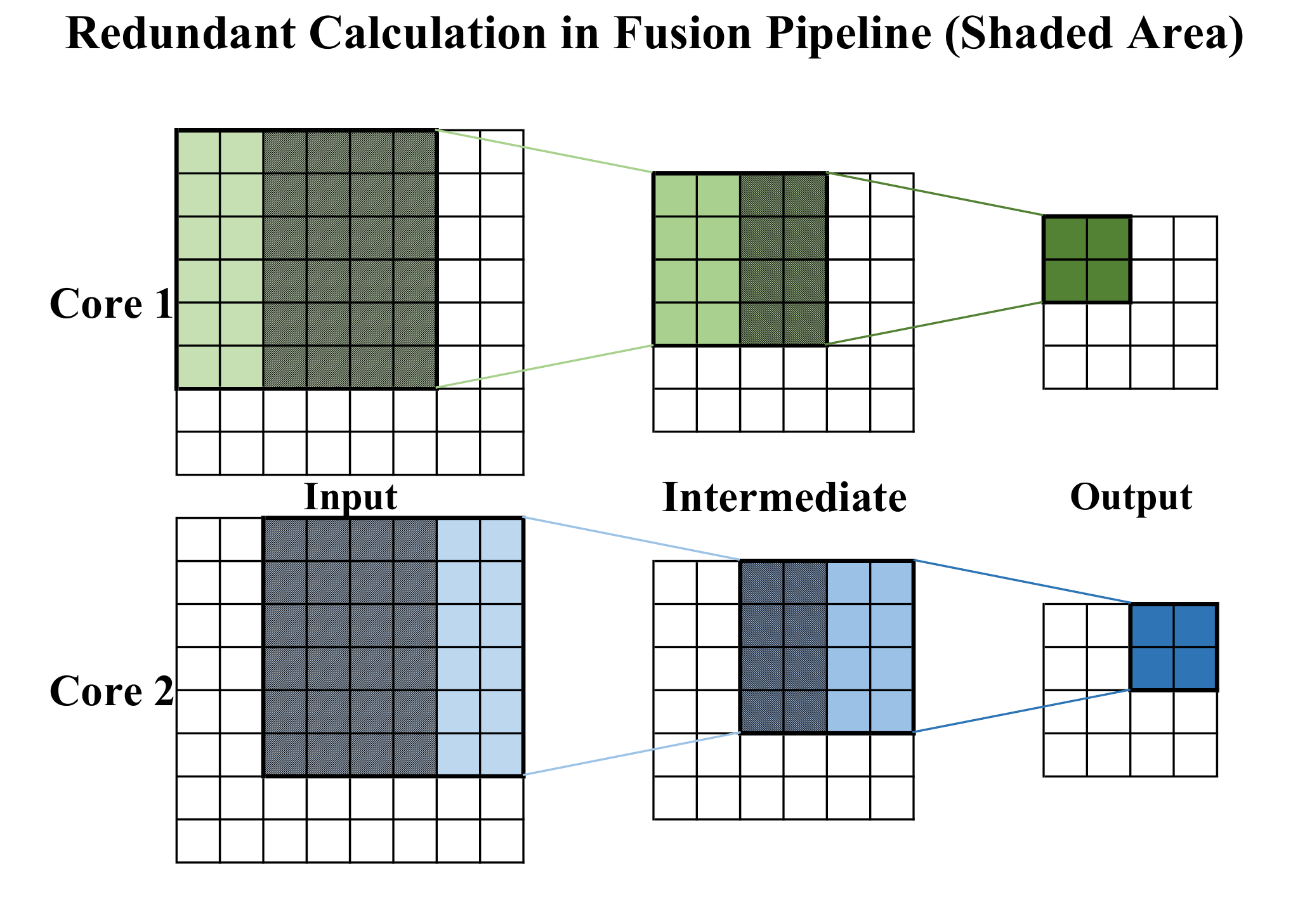}
	\includegraphics[width=0.32\linewidth]{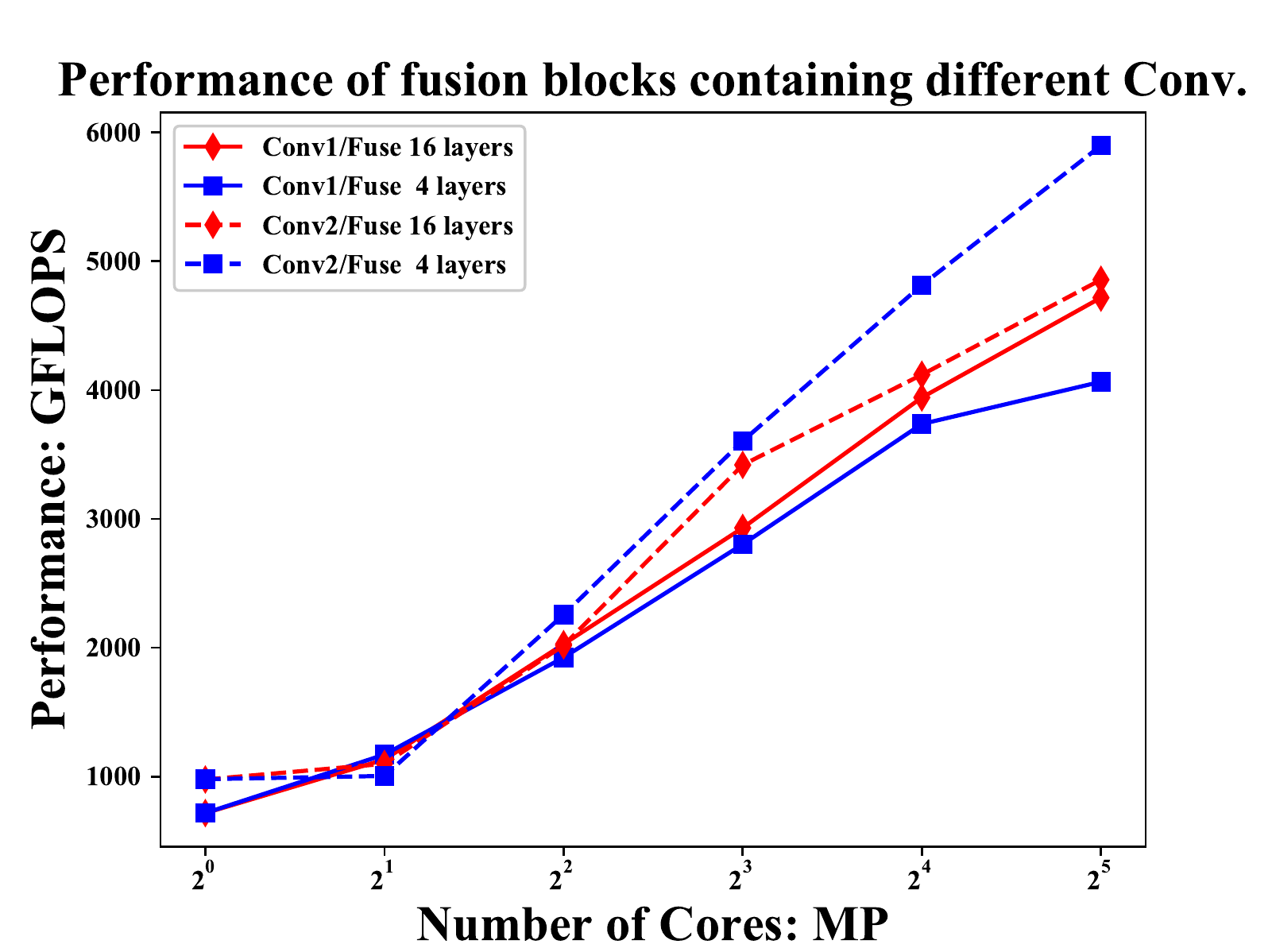}
	\includegraphics[width=0.32\linewidth]{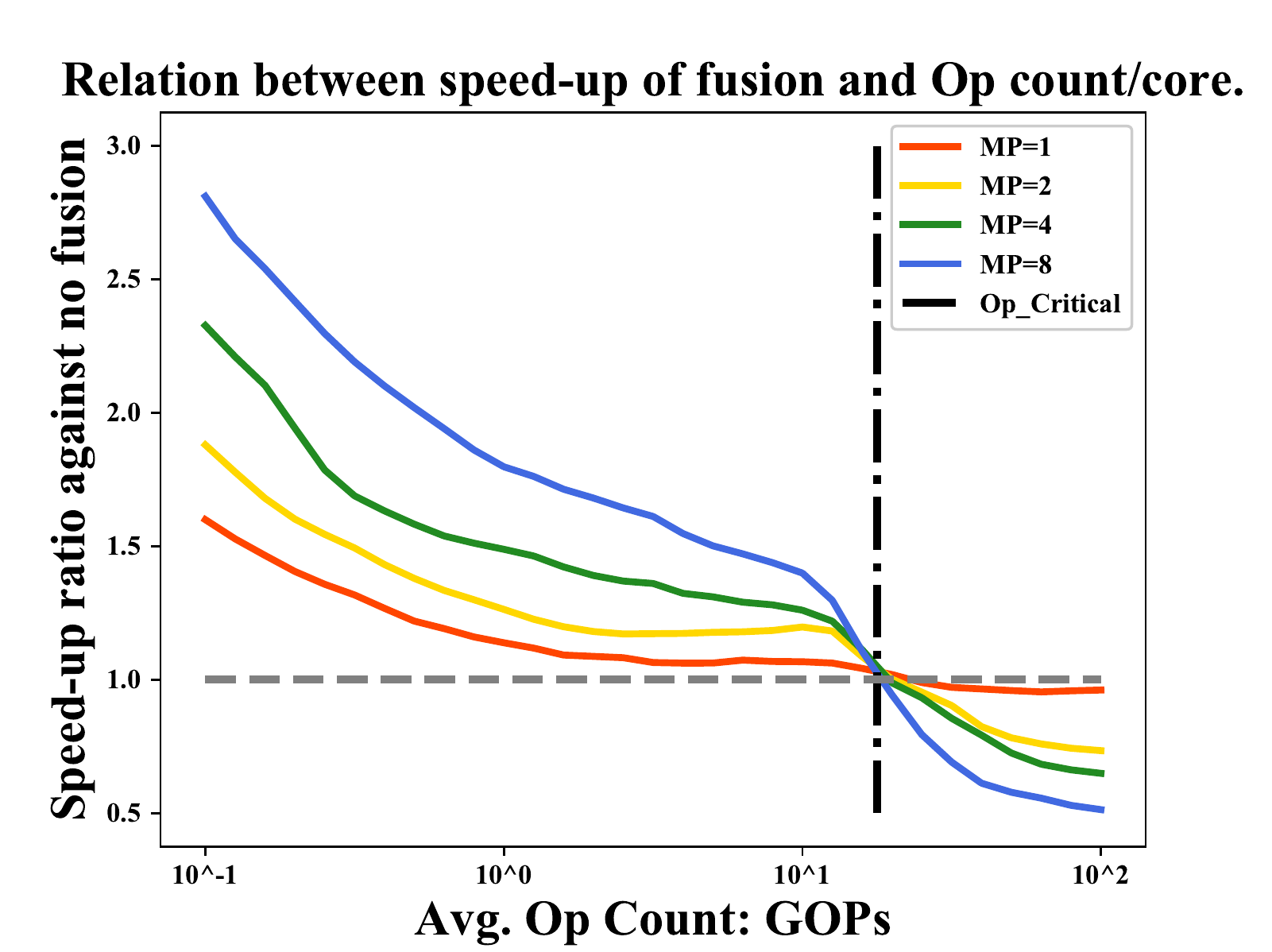}
	
	\caption{\label{fig-Fusion} (a) Redundant computation in layer fusion~\cite{FusedLayerCNN}. (b) Fusing different \texttt{CONV} layers. (c) Relation between speed-up ratio and the cores used.}
\end{figure*}
\begin{figure}[t]
	\centering
	\includegraphics[width=0.8\linewidth]{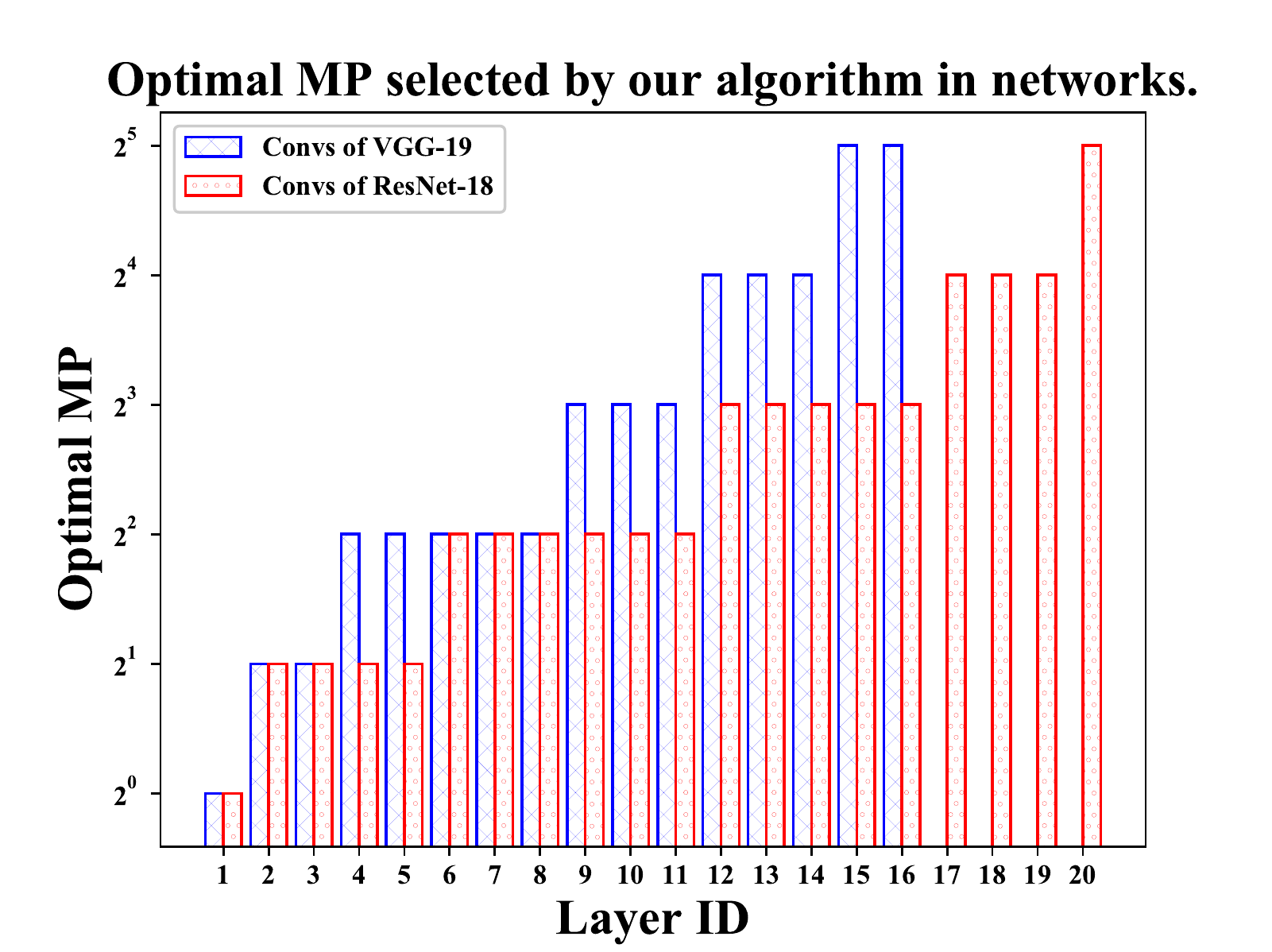}\\
	\includegraphics[width=0.8\linewidth]{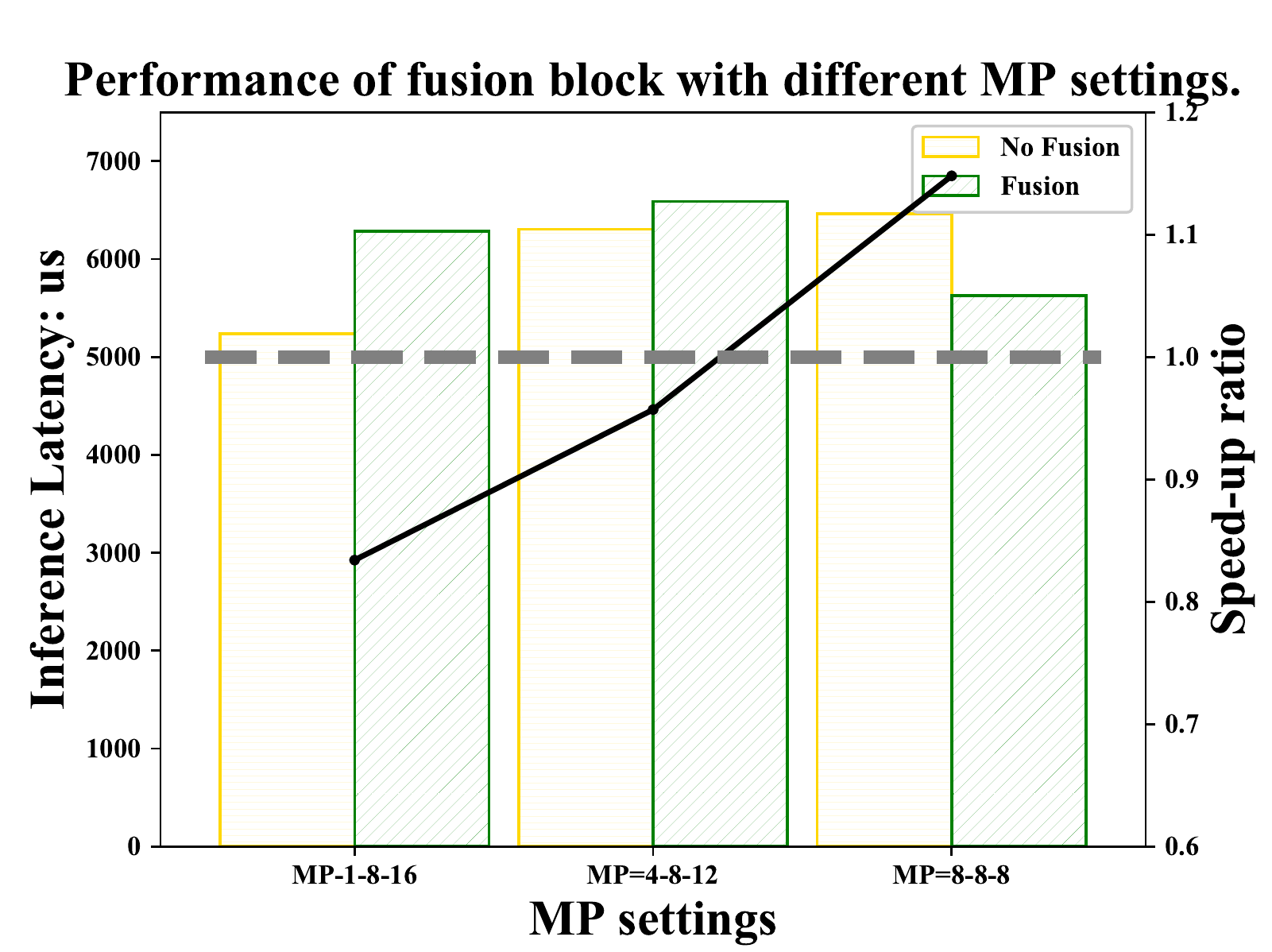}
	
	\caption{\label{fig-MPFusion} (a) Optimal MP selected by our method in ResNet-18~\cite{ResNet} and VGG-19~\cite{VGG}. (b) Performance and the speed-up ratio of fusion block containing convolutions with different optimal MP. Generally, we should determine the optimal MP given the convolution layer parameters. However, in this experiment, since we want to observe the influence of different optimal MP of convolutions in a fusion block, we determine the MP first and than determine the convolution parameters according to selected MP.}
\end{figure}

\subsection{Single Layer MP}

In \Sec{sec-BGD}, we show that the operation count of the \texttt{Conv} layer has a significant impact on its optimal \texttt{MP}.
However, operation count along can not be used for deciding the layer's optimal \texttt{MP}. 
In this subsection, we propose a joint model that uses both the operation count and in/out channel size of convolutional layers to determine the optimal \texttt{MP} configuration for the \texttt{Conv} layer.

We conduct a more detailed \texttt{MP} impact analysis for three different layers ($\{32, 32, 224\times224, 3\times3\}, \{64, 64, 112\times112, 3\times3\}, \{128, 128, 56\times56, 3\times3\}$).
In short, those three layers have the same operation count but different input/output channels.
\Fig{fig-SMCPerfDetail}(a) shows that those three layers have different optimal \texttt{MP} values.
This is because the hardware partitions the tensor on channel dimension with a certain minimal partition size.
In addition, we also find the operation count itself impacts the optimal \texttt{MP} values.
As \Fig{fig-SMCPerfDetail}(b) shows, \texttt{Conv} layers with the same input/output channel but different operation count have different optimal \texttt{MP} values.
\texttt{Conv} layers with fewer channels with high operation count could prefer more cores than layers with more channels but less operation count.
Given those findings, we use both a layer's channel size ($ C $ in the formula) and operation count to determine its optimal core number, as shown in Equation~\ref{equ-MP}, where $ \alpha,\beta $ are hardware-tuned scaling factors. We emperically decide the value of $\alpha$ and $\beta$ for MLU100 is 0.316 and 0.659 respectively according to the weight result of PCA.
\begin{equation}
\label{equ-MP}
MP(C, OpCount) \propto \alpha\times \log_{2}(C) + \beta\times \log_{2}(OpCount)
\end{equation}

\subsection{Multiple Layers Fusion and MP}
As presented in~\Sec{sec-BGD}, layer fusion can reduce the data movement and increase the operation count dispatched to cores to improve the performance. 
The fusion block composed of multiple layers can also leverage multiple cores to further reduce the latency.
On the other hand, using more mores for the fusion block also leads to redundant computation owing to the halo effect of 2D-convolution illustrated in \Fig{fig-Fusion}(a).
Moreover, the redundant computation increases when the number of layers in the fusion block grows, which means fusing more layers does not necessarily improve the performance.
To efficiently enable fusion into our optimization procedure, we first present our insight using several identical layers to illustrate the factors that influence the performance of fusion blocks. Then, given the significant layer heterogeneities in actual neural networks, we present our DLFusion algorithm for joint optimization considering both MP and fusion schemes. 

\subsubsection{Identical Layers}
We use two different \texttt{Conv} layers and compare their performance when fusing 4 and 16 layers.  
\Fig{fig-Fusion}(b) shows the performance comparison where \texttt{Conv1} is $\{ 512,512,28\times28,3\times3 \}$ and \texttt{Conv2} is $\{ 512,512,14\times14,3\times3 \}$, they have 1.72~GOPs and 0.43~GOPs respectively.
While fusing more layers for \texttt{Conv2} leads to better performance, the situation is opposite for \texttt{Conv1}.
The main reason is that when a large fusion block uses more cores, it leads to the more redundant computation. As shown in \Fig{fig-Fusion}(c), before reaching the critical operation count, using fusion can deliver better performance than not using fusion, since the single-core performance increase rapidly before the critical point according to \Fig{fig-rfline}(b). Once exceeding the critical point, the performance drops significantly due to the redundant computation (and that's why the single-core performance is stable before and after the critical point because using a single core will not introduce redundant computation). It should be noted that when using more cores, the critical value is slightly smaller also because of the redundant computation account for more op count. Moreover, before reaching the critical value, using more cores leads to better performance since more parallelism can be leveraged. For single-core, the improvement gained by fusion is mainly from reduced memory round-trip.
So, for layer fusion, we should limit the size of fusion block close to but below critical operation count of the cores to benefit the most from parallelism and avoid unacceptable redundant computation.

\subsubsection{Non-identical Layers}
In the actual DNN models, we find that the layers have different parameters that lead to different optimal MP as shown in \Fig{fig-MPFusion}(a). 
When fusing layers with significant different optimal MPs into one block, severe underperformance is observed as \Fig{fig-MPFusion}(b). The reason is that all layers in one fusion block will share the same MP, which is only suitable for a small part of layers in the block. Given this finding, we choose to determine optimal MP of every single layer first to avoid this underperformance for later joint optimization with fusion (since we can gather layers with the same or similar optimal MP together).
Once the optimal MPs are determined, to strike a balance between the increased operation count and redundant computation in layer fusion, we use the following heuristics. When performing layer fusion, we gradually fuse layers until the operation count of fused layers is greater than a preset threshold, which provides enough parallelism for the hardware and bounds the amount of redundant computation.

\subsection{Implementation}
In this subsection, we first present the core of our work: the DLFusion optimization algorithm, and then we introduce how we implement and evaluate our algorithm on actual hardware. 
\begin{algorithm*}[h]
	\caption{Finding fusion scheme and hyper-parameter setting}
	\label{Algo-Opt}
	\textbf{Input:}  onnx\_file, $num\_of\_layers$, $ OpCount_{critical} $\\
	\textbf{Output:}  $ fusion\_partition\_index[] $,  $ mp\_of\_fusionblock[] $
	\begin{algorithmic}[1]
		\Function {JointOptFusionAndMP}{onnx\_file, $ num\_of\_layer$, $ OpCount_{critical} $}
		\State $ layers\_spec \gets [], sum\_Op \gets 0 $
		\State $ current\_mp \gets 0, avg\_mp \gets 0, block\_size\gets 0 $
		\For {$i=0 \to num\_of\_layer $}
		\State $ layers\_spec[i] \gets $ Specification of $ i^{th} $ layer interpreted by TVM.Relay
		\If {$ layers\_spec[i].type = Convolution / Fully-Connected $}
		\State $ current\_mp \gets $ selection based on channel(major) and Op count(minor)
		\State $ sum\_Op \gets sum\_Op + $ operation count of $ i^{th} $ layer
		\State $ avg\_mp \gets avg\_mp + current\_mp, block\_size\gets block\_size + 1 $
		\EndIf
		\State $ avg\_mp \gets \frac{avg\_mp}{block\_size} $
		\If {$ \frac{sum\_Op}{avg\_mp} \geq OpCount_{critical} $}
		\State $ fusion\_partition\_index.push(i) $
		\State $ mp\_of\_fusionblock.push(2^{\lfloor \log_{2}(avg\_mp) \rfloor}) $
		\State $ sum\_Op \gets 0, avg\_mp\gets 0, block\_size\gets 0 $
		\EndIf
		\EndFor
		\State \Return $ fusion\_partition\_index, mp\_of\_fusionblock $
		\EndFunction
	\end{algorithmic}
\end{algorithm*}
\subsubsection{DLFusion Algorithm}
The DLFusion optimization algorithm aims to find a schedule with optimal MPs and fusion schemes for the hardware. We use the pseudo-code shown in Algorithm~\ref{Algo-Opt} for the joint optimization of fusion scheme and optimal \texttt{MP}.
Our algorithm requires the input of ONNX-based neural network description files, number of layers, and $OpCount_{critical}$, which is a tunable parameter that represents the operation count required by a single core to reach its peak performance. For MLU100, we choose this parameter as $ 10^{1.25} $GOPs as suggested in \Fig{fig-rfline}(b) and \Fig{fig-Fusion}(c).
The interpreter first reads the network parameters (Line 5).
The algorithm then decides the optimal \texttt{MP} for each \texttt{CONV} layer based on its channel dimension and operation count (Line 8).
The algorithm adds the current layer into the fusion block. It calculates the current total operation count and average \texttt{MP} for all the current fused layers (Line 8 to 11).
If the operation count dispatched to each core exceeds the critical operation count $OpCount_{critical}$, we stop the fusion for the current block and start a new fusion block (Line 12 to 13).
For the newly formed fusion block, we decide its \texttt{MP} as the closed to average \texttt{MP} and round it to $ 2^n $ (Line 14 to 15).
This process repeats until all layers are processed.
\subsubsection{DLFusion Compiler Tool Chain}
\begin{figure}[t]
	\centering
	\includegraphics[width=0.99\linewidth]{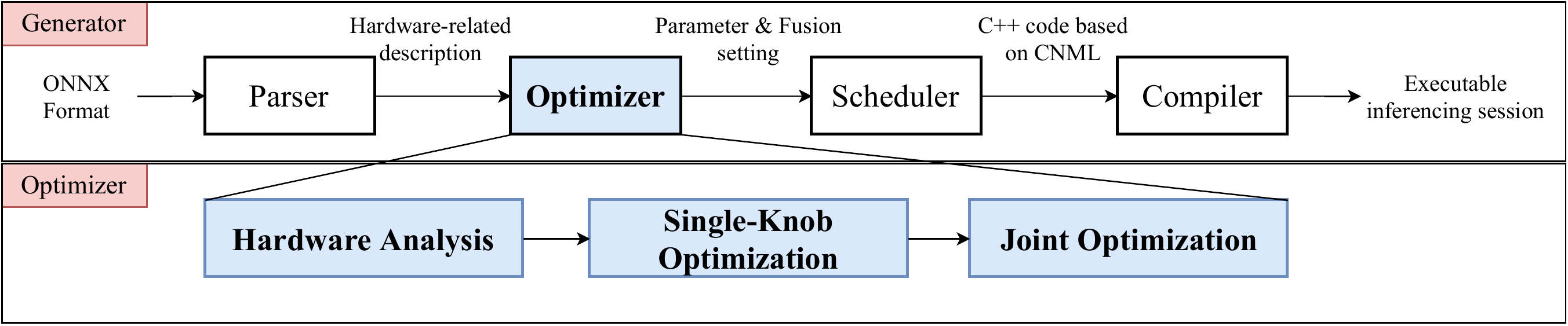}
	
	\caption{\label{fig-Flow} Overall architecture of our framework.}
\end{figure}

To evaluate the aforementioned optimization algorithm, we design and implement a compiler tool-chain for Cambricon MLU-100.
\Fig{fig-Flow} shows the details of our framework containing code generator and optimizer, which takes the input of ONNX format based neural network description file and generates the C++ code that leverages the MLU100's operator-level SDK CNML.
The core in the framework is the optimizer, which is a specialized instance of \Fig{fig-arch} and includes an optimization pass for tuning the execution parameters according to the DNN model characteristics.


The code generator produces the C++ source code based on a template file to call the CNML library.
The produced source code can be compiled to the executable inference session via the \emph{g++} compiler.
In our work,  we choose the DNN format ONNX~\cite{ONNX} because it is independent of specific deep learning frameworks.
We use the TVM~\cite{TVM} as the parser to convert the ONNX-based network description format to the TVM's internal graph representation that the following scheduler and optimizer use.

%% file: Evaluation.tex
\section{Evaluation}
In this section, we evaluate DLFusion using a set of representative CNN models, including ResNet, VGG, AlexNet and mobileNet, as listed in \Tbl{tab-Networks}.
We focus on the inference and use the frame per second (FPS) as the performance metric.
To demonstrate the effectiveness of DLFusion, we compare the performance of different optimization strategies, including a reduced brute-force search strategy.
\begin{table}[h]
	\caption{Networks description.}
	\label{tab-Networks}
	\centering
	\begin{tabular}{|c|c|c|c|}
		\hline
		Network & Total Op& Avg. Op & No. of \texttt{CONV}\\
		\hline
		ResNet-18/50~\cite{ResNet} & 3.38/7.61 & 0.169/0.144 & 20/53 \\ \hline
		VGG-19~\cite{VGG} & 36.34 & 2.27 & 16 \\ \hline
		AlexNet~\cite{AlexNet} & 1.22 & 0.244 & 5 \\ \hline
		mobileNet~\cite{mobileNet} & 10.33 & 0.199 & 52 \\ 
		\hline
	\end{tabular}
\end{table}

\newcommand{\tabincell}[2]{\begin{tabular}{@{}#1@{}}#2\end{tabular}}
\begin{table*}[h]
	\caption{Different optimization strategies.}
	\label{tab-OPT}
	\centering
	\begin{tabular}{|c|c|c|}
		\hline
		Index & Strategy Name & Description  \\
		\hline
1 & Non-Optimization 	& No fusion with $MP=1$\\ \hline
2 & Fixed MP 	& No fusion: all the layers have the same MP.\\ \hline
3 & Dynamic MP		&  No fusion: each layer its own MP. \\ \hline
4 & All Fusion \& Max. MP & \tabincell{c}{All the layers are fused into one block, \\ and MP is set to be 32 (maximal)}\\ \hline
5 & Fusion \& Fixed MP & \tabincell{c}{Fuse layers to multiple blocks (Alg.~\ref{Algo-Opt}), \\and all blocks have the same MP.}\\ \hline
6 & \tabincell{c}{\textbf{DLFusion}\\(Fusion \& Dynamic MP)}  & \tabincell{c}{Use Alg.~\ref{Algo-Opt} to fuse layers and \\ set MP for each fused block.}\\ \hline
7 & Brute-force Search & Optimal performance.\\
		\hline
	\end{tabular}
\end{table*}

\begin{figure*}[t]
	\centering
	\includegraphics[width=0.99\linewidth]{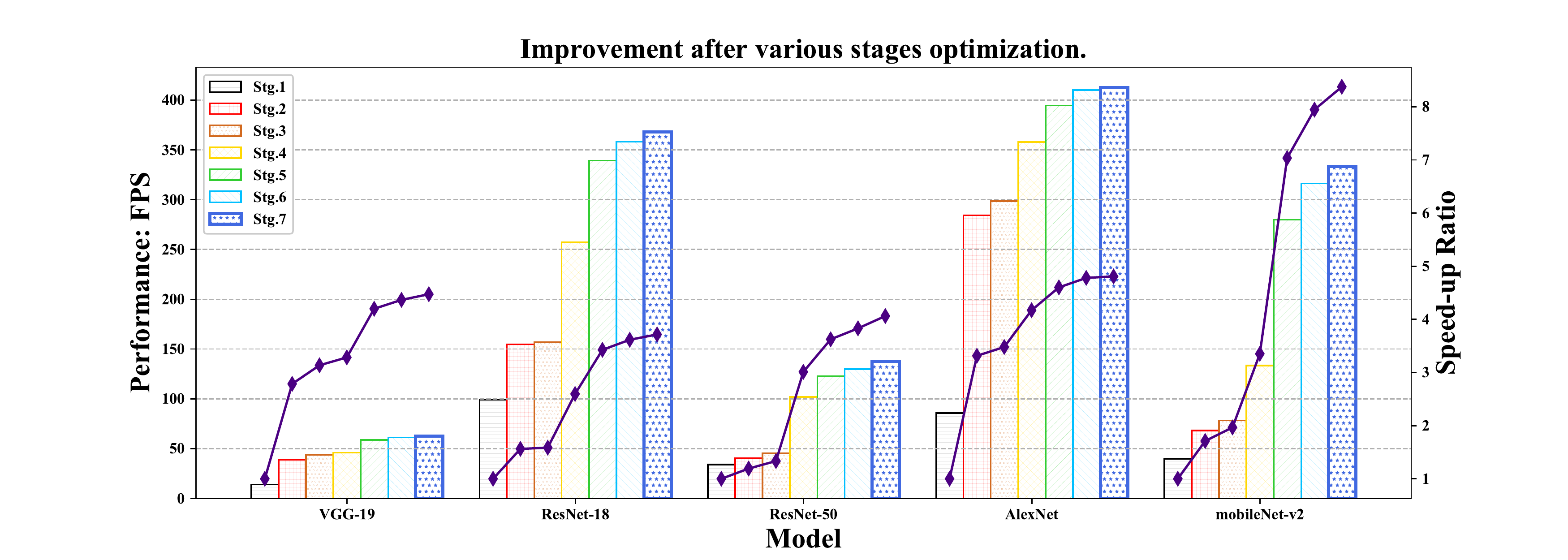}
	\caption{\label{fig-Improvement} Performance comparison of different optimization strategies.}

\end{figure*}

\subsubsection{Evaluated Strategy}
We evaluate different optimization strategies that are listed in \Tbl{tab-OPT}.
Strategy 1 referring to no fusion and no model parallelism ($MP = 1$) is used as the baseline.
Both strategy 2 and 3 perform no fusion while the former uses the same $MP$ for all layers, and the latter uses the layer-specific $MP$.
Strategy 4 simply fuses all layers and uses the maximal $MP$ for the fused block.
Strategy 5 uses Alg.~\ref{Algo-Opt} to fuse layers to multiple blocks and use a single fixed $MP$ for all blocks.
In contrast, Strategy 6  uses Alg.~\ref{Algo-Opt} to fuse layers and set a block-specific $MP$ value for each block.
Strategy 7 represents the optimal performance through a brute-force search that we detail later.


\subsubsection{Performance Comparison}
\Fig{fig-Improvement} shows the performance comparison of different optimization strategies.
Except for the oracle case (the last bar), DLFusion has the best performance, which achieves a speedup of $ 3.6$ - $ 7.9\times $ against the baseline. 
Our algorithm leads to significant performance improvement for the two following reasons. 
First, each fusion block has a proper operation count that gains plenty of parallelism while with acceptable redundant computation. 
Second, the number of cores used in each fusion block is also close to their optimal number of cores that balances computation and memory access. 
The studied CNN models have different performance trend over the different optimization strategies, for which we make the following observations.
\begin{itemize}[label=$\bullet$, leftmargin=*]
	\item CNN models with low operation count per layer (e.g., ResNet and mobileNet) are not sensitive to \texttt{MP} optimization because using more cores leads to less utilization of each core. In contrast, the model with high operation count per layers (e.g., VGG-19) benefits more from \texttt{MP} optimization.
	\item CNN models with low operation count per layer (e.g., ResNet and mobileNet) benefits significantly from the layer fusion optimization because layer fusion produces a block with more operations, which results in performance improvement. In contrast, the model with high operation count per layers (e.g., VGG-19) benefits less from layer fusion. 
	\item With the increasing of the number of layers (from 4 \texttt{CONV} layers in AlexNet to over 50 \texttt{CONV} layers in mobileNet and ResNet-50), the model gets more sensitive to the fusion strategy because of larger fusion scheme space.
\end{itemize}

\subsubsection{Oracle Case}
To evaluate the effectiveness of the DLFusion approach, we design a feasible brute-force search as the oracle case.
As we explain in \Sec{sec:motivation}, the hyper-parameter space is too large for the brute-force search.
We reduce the search space based on the performance characterization analysis on the existing CNN models.
First, we limit the choice of MP from $ {1,2,3...32} $ to $ {1,2,4,8,12,16,24,32} $. Second, we limit the size of a fusion block to the multiple of four. 
These two rules lead to acceptable search time.
The last column of each CNN model in \Fig{fig-Improvement} represents the oracle case achieved by our reduced brute-force search. 
The performance between the DLFusion and the oracle case is less than 10\%.
Meanwhile, with the increased number of layers, this performance gap gets smaller.
In general, DLFusion achieves the performance that is close to the oracle case, with much reduced search time ($O(n)$, where n is the number of layers).


%% file: related.tex
\section{Related Work}

To address the difficulty in compiler tool-chain design and optimization, researchers have proposed Domain Specific Language (DSL) to schedule the hardware more efficiently.
Halide is a popular DSL for image processing pipelines~\cite{Halide}, and Taichi is proposed for CG processing~\cite{TaiChi}.
DSLs are concise by omitting control logic in the regular programming language, which makes it more convenient for optimization.
Given the similarity between image processing pipeline and DNN models, Halide-based compilation frameworks such as TensorComprehension~\cite{TC}  and TVM~\cite{TVM}  have been proposed.
Those frameworks target a general optimization at the computation graph level.
In contrast, DLFusion is a hardware-specific optimization framework  that can be integrated as a backend for the graph-level frameworks.

Previous researchers have studied various general optimizations, such as loop fusion~\cite{LoopKernelFusion} and kernel fusion~\cite{KernelFusion,DSL}. 
In the TensorComprehension~\cite{TC} framework, fusion is performed with the use of the polyhedral model~\cite{Polyhedral}. 
TASO~\cite{TASO} explores the graph substitution optimization using a cost-based backtracking search. Grappler~\cite{Grappler} of TensorFlow conducts a series of rule-based arithmetic transformation, operation fusion. On the other hand, instead of optimize the execution on single hardware, Google REGAL~\cite{REGAL} focus on the problem of scheduling the execution on multiple hardware. 
Those optimizations work at the graph-level and can be used for any hardware back-end.

Other possible optimization options for compilers include batching~\cite{8988602}, model sparsity~\cite{CambriconS,TileWiseSparsity}, and data movement reduction between specialized hardware accelerators and general purpose processors~\cite{SMA}.
Besides performance optimization, increasing efforts have been put on the robustness of the DNN systems including Ptolemy, an architecture that detect adversarial samples at inference time~\cite{Ptolemy} based on critical path method~\cite{AdversarialDefense}. Other researchers have also explored traditional reliability on heterogeneous systems~\cite{Asymmetric}. 
To generalize the compiling stage optimization to these architectures is also an urgent need.
DLFusion targets the layer fusion optimization that is specific to DNN models on a specific hardware accelerator, researchers have also explored the programming framework of this accelerator that support the optimization options with less code effort~\cite{Paleozoic}. Prior work explores the layer fusion as an architecture optimization~\cite{FusedLayerCNN} while we use it for compiler optimization.






%% file: Conclusion.tex
\section{Conclusion}

In this work, we propose an end-to-end code generator with optimizer for the DNN accelerator Cambricon-MLU100, which is capable of generating optimized C++ code for a DNN model with ONNX format.
We propose an auto-tuning algorithm to jointly optimize the two execution hyper-parameter (i.e., number of cores and layer fusion scheme) to maximize the accelerator performance for a given DNN model.
The algorithm uses the operation count and channel size as the features to decide the optimal core count for each layer.
It then gradually fuses layers into a block that has just enough computation to fully utilize the hardware and avoids excessive redundant computation.
Evaluation shows that our approach achieves almost the same performance of the reduced brute-force search base oracle case, but with a much less search time.
To the best of our knowledge, our work represents the first auto-tuning algorithm for a DNN accelerator can we hope it can foster more research efforts in this direction.



%% file: Reference.tex
\printbibliography

%% file: ref.bib
@inproceedings{Halide,
  author    = {J. Ragan{-}Kelley and
               Connelly Barnes and
               Andrew Adams and
               Sylvain Paris and
               Fr{\'{e}}do Durand and
               Saman P. Amarasinghe},
  title     = {Halide: a language and compiler for optimizing parallelism, locality,
               and recomputation in image processing pipelines},
  booktitle = {Conference on Programming Language Design and Implementation},
  year      = {2013}
}

@inproceedings{AdversarialDefense,
  author    = {Yuxian Qiu and
               Jingwen Leng and
               Cong Guo and
               Quan Chen and
               Chao Li and
               Minyi Guo and
               Yuhao Zhu},
  title     = {Adversarial Defense Through Network Profiling Based Path Extraction},
  booktitle = {{IEEE} Conference on Computer Vision and Pattern Recognition, {CVPR}
               2019, Long Beach, CA, USA, June 16-20, 2019},
  pages     = {4777--4786},
  publisher = {Computer Vision Foundation / {IEEE}},
  year      = {2019}
}

@inproceedings{Asymmetric,
  author    = {Jingwen Leng and
               Alper Buyuktosunoglu and
               Ramon Bertran and
               Pradip Bose and
               Quan Chen and
               Minyi Guo and
               Vijay Janapa Reddi},
  title     = {Asymmetric Resilience: Exploiting Task-Level Idempotency for Transient
               Error Recovery in Accelerator-Based Systems},
  booktitle = {{IEEE} International Symposium on High Performance Computer Architecture,
               {HPCA} 2020, San Diego, CA, USA, February 22-26, 2020},
  pages     = {44--57},
  publisher = {{IEEE}},
  year      = {2020}
}

@inproceedings{SMA,
  author    = {Cong Guo and
               Yangjie Zhou and
               Jingwen Leng and
               Yuhao Zhu and
               Zidong Du and
               Quan Chen and
               Chao Li and
               Bin Yao and
               Minyi Guo},
  title     = {Balancing Efficiency and Flexibility for {DNN} Acceleration via Temporal
               GPU-Systolic Array Integration},
  booktitle = {57th {ACM/IEEE} Design Automation Conference, {DAC} 2020, San Francisco,
               CA, USA, July 20-24, 2020},
  pages     = {1--6},
  publisher = {{IEEE}},
  year      = {2020}
}

@inproceedings{CambriconS,
  author    = {Xuda Zhou and
               Zidong Du and
               Qi Guo and
               Shaoli Liu and
               Chengsi Liu and
               Chao Wang and
               Xuehai Zhou and
               Ling Li and
               Tianshi Chen and
               Yunji Chen},
  title     = {Cambricon-S: Addressing Irregularity in Sparse Neural Networks through
               {A} Cooperative Software/Hardware Approach},
  booktitle = {51st Annual {IEEE/ACM} International Symposium on Microarchitecture,
               {MICRO} 2018, Fukuoka, Japan, October 20-24, 2018},
  pages     = {15--28},
  publisher = {{IEEE} Computer Society},
  year      = {2018}
}

@article{TileWiseSparsity,
  author    = {Cong Guo and
               Bo Yang Hsueh and
               Jingwen Leng and
               Yuxian Qiu and
               Yue Guan and
               Zehuan Wang and
               Xiaoying Jia and
               Xipeng Li and
               Minyi Guo and
               Yuhao Zhu},
  title     = {Accelerating Sparse {DNN} Models without Hardware-Support via Tile-Wise
               Sparsity},
  journal   = {CoRR},
  volume    = {abs/2008.13006},
  year      = {2020}
}

@inproceedings{VGG,
	author    = {Karen Simonyan and
	Andrew Zisserman},
	title     = {VVGGery Deep Convolutional Networks for Large-Scale Image Recognition},
	booktitle = {3rd International Conference on Learning Representations},
	year      = {2015}
}

@article{Ptolemy,
  author    = {Yiming Gan and
               Yuxian Qiu and
               Jingwen Leng and
               Minyi Guo and
               Yuhao Zhu},
  title     = {Ptolemy: Architecture Support for Robust Deep Learning},
  journal   = {CoRR},
  volume    = {abs/2008.09954},
  year      = {2020}
}

@article{TaiChi,
  author    = {Yuanming Hu and
               Tzu{-}Mao Li and
               Luke Anderson and
               Jonathan Ragan{-}Kelley and
               Fr{\'{e}}do Durand},
  title     = {Taichi: a language for high-performance computation on spatially sparse
               data structures},
  journal   = {{ACM} Trans. Graph.},
  volume    = {38},
  number    = {6},
  pages     = {201:1--201:16},
  year      = {2019}
}

@article{TensorFlowXLA,
  author    = {Ruben Mayer and
               Christian Mayer and
               Larissa Laich},
  title     = {The TensorFlow Partitioning and Scheduling Problem: It's the Critical
               Path!},
  volume    = {abs/1711.01912},
  year      = {2017}
}

@article{roofline,
	author    = {Samuel Williams and
	Andrew Waterman and
	David A. Patterson},
	title     = {Roofline: an insightful visual performance model for multicore architectures},
	journal   = {Commun. {ACM}},
	volume    = {52},
	number    = {4},
	pages     = {65--76},
	year      = {2009},
	doi       = {10.1145/1498765.1498785}
}

@inproceedings{TVM,
	author    = {Tianqi Chen and
	Thierry Moreau and
	Ziheng Jiang and
	Lianmin Zheng and
	Eddie Q. Yan and
	Haichen Shen and
	Meghan Cowan and
	Leyuan Wang and
	Yuwei Hu and
	Luis Ceze and
	Carlos Guestrin and
	Arvind Krishnamurthy},
	title     = {{TVM:} An Automated End-to-End Optimizing Compiler for Deep Learning},
	booktitle = {13th {USENIX} Symposium on Operating Systems Design and Implementation},
	year      = {2018}
}

@article{TC,
	author    = {N. Vasilache and
	Oleksandr Zinenko and
	Theodoros Theodoridis and
	Priya Goyal and
	Zachary DeVito and
	William S. Moses and
	Sven Verdoolaege and
	Andrew Adams and
	Albert Cohen},
	title     = {Tensor Comprehensions: Framework-Agnostic High-Performance Machine
	Learning Abstractions},
	journal   = {CoRR},
	volume    = {1802.04730},
	year      = {2018}
}

@inproceedings{DSL,
	author    = {Bo Qiao and
	Oliver Reiche and
	Frank Hannig and
	J{\"{u}}rgen Teich},
	title     = {Automatic Kernel Fusion for Image Processing DSLs},
	booktitle = {Proceedings of the 21st International Workshop on Software and Compilers
	for Embedded Systems},
	year      = {2018}
}

@inproceedings{KernelFusion,
	author    = {Guibin Wang and
	Yisong Lin and
	Wei Yi},
	title     = {Kernel Fusion: An Effective Method for Better Power Efficiency on
	Multithreaded {GPU}},
	booktitle = {2010 {IEEE/ACM} Int'l Conference on Green Computing and Communications},
	year      = {2010}
}

@article{Paleozoic,
    author = {Zihan Liu and Jingwen Leng and Guandong Lu and Quan Chen and Chenhui Wang and Minyi Guo},
    title = {Survey and design of paleozoic: a high-performance compiler tool chain for deep learning inference accelerator},
    journal = {CCF Trans. of High Performance Computing},
    year = {2020}
}

@inproceedings{PuDianNao,
	author    = {Dao{-}Fu Liu and
	Tianshi Chen and
	Shaoli Liu and
	Jinhong Zhou and
	Shengyuan Zhou and
	Olivier Temam and
	Xiaobing Feng and
	Xuehai Zhou and
	Yunji Chen},
	title     = {PuDianNao: {A} Polyvalent Machine Learning Accelerator},
	booktitle = {Proceedings of the International Conference on Architectural
	Support for Programming Languages and Operating Systems},
	year      = {2015}
}

@inproceedings{LoopKernelFusion,
	author    = {Bo Qiao and
	Oliver Reiche and
	Frank Hannig and
	J{\"{u}}rgen Teich},
	title     = {From Loop Fusion to Kernel Fusion: {A} Domain-Specific Approach to
	Locality Optimization},
	booktitle = {{IEEE/ACM} International Symposium on Code Generation and Optimization},
	year      = {2019}
}

@inproceedings{TASO,
	author    = {Zhihao Jia and
	Oded Padon and
	James J. Thomas and
	Todd Warszawski and
	Matei Zaharia and
	Alex Aiken},
	title     = {{TASO:} optimizing deep learning computation with automatic generation
	of graph substitutions},
	booktitle = {Proceedings of the 27th {ACM} Symposium on Operating Systems Principles},
	year      = {2019}
}

@inproceedings{Polyhedral,
	author    = {Uday Bondhugula and
	Albert Hartono and
	J. Ramanujam and
	P. Sadayappan},
	title     = {A practical automatic polyhedral parallelizer and locality optimizer},
	booktitle = {Proceedings of the Conference on Programming
	Language Design and Implementation},
	year      = {2008}
}

@inproceedings{TPU,
	author    = {Norman P. Jouppi and
	Cliff Young and
	Nishant Patil and
	David A. Patterson and
	Gaurav Agrawal and
	Raminder Bajwa and
	Sarah Bates and
	Suresh Bhatia and
	Nan Boden and
	Al Borchers and
	Rick Boyle and
	Pierre{-}luc Cantin and
	Clifford Chao and
	Chris Clark and
	Jeremy Coriell and
	Mike Daley and
	Matt Dau and
	Jeffrey Dean and
	Ben Gelb and
	Tara Vazir Ghaemmaghami and
	Rajendra Gottipati and
	William Gulland and
	Robert Hagmann and
	C. Richard Ho and
	Doug Hogberg and
	John Hu and
	Robert Hundt and
	Dan Hurt and
	Julian Ibarz and
	Aaron Jaffey and
	Alek Jaworski and
	Alexander Kaplan and
	Harshit Khaitan and
	Daniel Killebrew and
	Andy Koch and
	Naveen Kumar and
	Steve Lacy and
	James Laudon and
	James Law and
	Diemthu Le and
	Chris Leary and
	Zhuyuan Liu and
	Kyle Lucke and
	Alan Lundin and
	Gordon MacKean and
	Adriana Maggiore and
	Maire Mahony and
	Kieran Miller and
	Rahul Nagarajan and
	Ravi Narayanaswami and
	Ray Ni and
	Kathy Nix and
	Thomas Norrie and
	Mark Omernick and
	Narayana Penukonda and
	Andy Phelps and
	Jonathan Ross and
	Matt Ross and
	Amir Salek and
	Emad Samadiani and
	Chris Severn and
	Gregory Sizikov and
	Matthew Snelham and
	Jed Souter and
	Dan Steinberg and
	Andy Swing and
	Mercedes Tan and
	Gregory Thorson and
	Bo Tian and
	Horia Toma and
	Erick Tuttle and
	Vijay Vasudevan and
	Richard Walter and
	Walter Wang and
	Eric Wilcox and
	Doe Hyun Yoon},
	title     = {In-Datacenter Performance Analysis of a Tensor Processing Unit},
	booktitle = {Proceedings of the 44th Annual International Symposium on Computer
	Architecture},
	year      = {2017}
}

@inproceedings{REGAL,
	author    = {Aditya Paliwal and
	Felix Gimeno and
	Vinod Nair and
	Yujia Li and
	Miles Lubin and
	Pushmeet Kohli and
	Oriol Vinyals},
	title     = {Reinforced Genetic Algorithm Learning for Optimizing Computation Graphs},
	booktitle = {8th International Conference on Learning Representations, {ICLR} 2020,
	Addis Ababa, Ethiopia, April 26-30, 2020},
	year      = {2020}
}

@inproceedings{AlexNet,
	author    = {A. Krizhevsky and others},
	title     = {ImageNet Classification with Deep Convolutional Neural Networks},
	booktitle = {Adv. in Neural Info. Proc. Systems},
	year      = {2012}
}

@misc{ONNX,
	howpublished = {\url{http://onnx.ai}},
	note = {Accessed Jun. 29, 2020},
	title = {Open Neural Network Exchange},
	subtitle = {The open standard for machine learning interoperability},
	author = {ONNX}
}

@misc{Grappler,
	howpublished = {\url{https://www.tensorflow.org/guide/graph_optimization}},
	note = {Access Jun. 29, 2020},
	title = {TensorFlow graph optimization with Grappler},
	author = {TensorFlow}
}

@inproceedings{mobileNet,
	author    = {M. Sandler and
	Andrew G. Howard and
	Menglong Zhu and
	Andrey Zhmoginov and
	Liang{-}Chieh Chen},
	title     = {MobileNetV2: Inverted Residuals and Linear Bottlenecks},
	booktitle = {Conference on Computer Vision and Pattern Recognition},
	year      = {2018}
}

@INPROCEEDINGS{8988602,
  author={W. {Cui} and M. {Wei} and Q. {Chen} and X. {Tang} and J. {Leng} and L. {Li} and M. {Guo}},
  booktitle={2019 IEEE 37th International Conference on Computer Design (ICCD)}, 
  title={Ebird: Elastic Batch for Improving Responsiveness and Throughput of Deep Learning Services}, 
  year={2019},
  volume={},
  number={},
  pages={497-505},
  doi={10.1109/ICCD46524.2019.00075}}

@manual{V100,
	author = {NVIDIA},
	title = {Tesla V100 Performance Guide},
	year = {2018}

}

@manual{TensorRT,
    author = {NVIDIA},
    title = {NVIDIA TensorRT: Programmable Inference Accelerator},
    year = {2020}
}

@manual{CambriconMLU100,
	author	=	{Cambricon Technologies},
	title = {Cambricon MLU100 Datasheet},
	month = {8},
	year = {2019}
}

@inproceedings{ResNet,
	author    = {Kaiming He and
	Xiangyu Zhang and
	Shaoqing Ren and
	Jian Sun},
	title     = {Deep Residual Learning for Image Recognition},
	booktitle = {2016 {IEEE} Conference on Computer Vision and Pattern Recognition},
	year      = {2016}
}

@inproceedings{FusedLayerCNN,
	author    = {Manoj Alwani and
	Han Chen and
	Michael Ferdman and
	Peter Milder},
	title     = {Fused-layer {CNN} accelerators},
	booktitle = {49th Annual {IEEE/ACM} International Symposium on Microarchitecture},
	year      = {2016}
}
